# Six-Dimensional Matching of Intense Beam with Linear Accelerating Structure

Yuri K. Batygin, Los Alamos National Laboratory, Los Alamos, NM 87545, USA

*Abstract*

Beam matching is a common technique that is routinely employed in accelerator design with the aim of minimizing beam losses and preservation of beam brightness. Despite being widely used, a full theoretical understanding of beam matching in 6D remains elusive. Here, we present an analytical treatment of 6D beam matching of a high-intensity beam onto an RF structure. We begin our analysis within the framework of a linear model, and apply the averaging method to a set of 3D beam envelope equations. Accordingly, we obtain a matched solution that is comprised of smoothed envelopes and periodic terms, describing envelope oscillations with the period of the focusing structure. We then consider the nonlinear regime, where the beam size is comparable with the separatrix size. Stating with a Hamiltonian analysis in 6D phase space, we attain a self-consistent beam profile and show that it is significantly different from the commonly used ellipsoidal shape. Subsequently, we analyze the special case of an equilibrium with equal space charge depression between all degrees of freedom. Comparison of beam dynamics for equipartitioned, equal space charge depression, and equal emittances beams is given. Finally, we present experimental results on beam matching in the LANSCE linac.

## 1. Introduction

The last two decades have seen significant advances in the construction and operation of a new generation of high-power particle accelerators, while new projects with more powerful beams are currently under consideration and construction [1-12]. However, the sustained and long-term operation of such machines is not without challenges: a key problem standing in the way of increased beam power is represented by beam losses associated with the expansion of phase space volume of the beam, and generation of a beam halo. In particular, beam losses in a linear accelerator result in radio-activation of the structure itself and deterioration of its components. In turn, this degrades the linac's reliability and hinders hands-on maintenance, limiting components lifespan.

The beam loss problem is broadly appreciated, and among the main practical approaches toward mitigating losses in real accelerators is a process known as beam matching. Qualitatively, beam matching refers to identification of a periodic solution for beam envelopes in periodic accelerator structures. While four-dimensional matching within a periodic focusing transport is a well-known procedure [13, 14], matching of a bunched beam in six-dimensional phase space is significantly less well - understood from a theoretical basis. More specifically, in case of weak space charge forces, matched beam Twiss parameters coincide with those of the accelerator lattice. In case of significant space charge, however, 6D matched beam parameters are different from lattice Twiss parameters and typically can only be determined numerically [15-17]. To this end, the related problem of deviation away from matched 3D beam envelopes and associated resonance envelope behavior has been explored in Refs. [18-20].

The traditional approach to six-dimensional beam matching is based on keeping transverse and longitudinal focusing strengths of lattice unchanged at transitions between accelerating sections [21]. The accelerator lattice in this case is a continuous focusing structure that changes adiabatically along the machine. This allows the beam - originally matched within the RFQ - to continue to be approximately matched along the linac, and to remain nearly independent of the space charge and emittance of the beam. Such an approach puts consequential constraints on lattice design. Particularly, discontinuities between accelerating sections must be sufficiently small to avoid significant mismatch with variation of beam parameters.

It is important to note that in real machines, discontinuities of this sort are practically unavoidable, since additional equipment must be routinely introduced between accelerating sections, such as choppers, beam collimators, deflectors, and beam diagnostics. To circumvent this issue, Medium Energy Beam Transports and other transitions are necessary to provide exact beam matching between accelerating sections. Transverse matching is achieved with the combination of quadrupoles, while longitudinal matching is accomplished using RF cavities [22-24]. By and large, the determination of matching conditions relies on numerical integration of particle dynamics in presence of space charge forces, and state-of-the-art codes such as TRACE-3D [25] and Trace-Win [26] have been widely employed to perform the matching. Let us note, that beam optics codes without inclusion of space charge forces, like TRANSPORT [27] and MAD [28], cannot be used for matching of intense beams, because matching conditions depend on beam current.

In contrast with the aforementioned advances in quantifying beam matching from computational and experimental grounds, a theoretical understanding of this process remains rudimentary by comparison. In contrast with the well-known KV distribution for a continuous, uniformly charged beam with elliptical cross-section in transport channels, there is no known solution for a self-consistent distribution function in 6D phase space [29-31]. Although widely used 3D envelope equations for a uniformly charge beam ellipsoid present a reasonable approximation to particle dynamics in an RF field [32], the collective theoretical understanding of beam matching in 6D falls short of a complete and comprehensive model. Addressing this shortcoming is the principal goal of our analysis. In other words, the primary purpose of this work is a purely analytical description of a matched beam solution, based upon 6D Hamiltonian and 3D beam envelope equations. To be clear, the Hamiltonian formalism presented in this work is by no means a replacement for the aforementioned numerical methods. Instead, the key aim of our analysis lies in presenting a complementary viewpoint that can illuminate the physical mechanisms underlying beam matching in modern high-intensity linacs.

Organization of the paper is as follows. Our analysis starts in Section 2 with a solution of 3D beam envelope equations using the averaging method, followed by a derivation of matched conditions for beam envelopes in Section 3. In Section 4, a new class of beam equilibria with equal space charge depression in transverse and longitudinal directions is discussed. In Section 5, an approximate self-consistent solution based on the Hamiltonian formalism in 6D phase space is applied to beam equilibrium, which yields a bunch shape notably different from the commonly used elliptical shape. Section 6 describes the results of simulations of a beam with three different equilibria: equal space-charge depression in transverse and longitudinal directions, equipartitioning, and equal beam emittances. Section 7 describes experimental results of beam matching in the LANSCE linear accelerator. The results are briefly summarized in Section 8. Appendixes A and B contain derivations of a 6D Hamiltonian for particle motion in RF as well as focusing fields and analysis of dynamics around synchronous particle.

## 2. Three-Dimensional Beam Envelope Equations

As particles experience acceleration, they perform nonlinear oscillations around a guiding synchronous particle. Such oscillations are described by a 6D Hamiltonian, which is presented in Appendix A. In many cases particle bunches are small enough for particle motion to be considered as that occurring in the vicinity of synchronous particle, where RF fields are closely approximated by linear functions of coordinates. Assuming that the bunched beam is uniformly charged ellipsoid, space charge forces are also linear functions of coordinates. Under these assumptions, equations of motion are linear and beam can be described by a set of 3D envelope equations. Analysis of linear dynamics around a synchronous particle is given in Appendix B.

Beam envelope equations are valid for non-uniform beams as well. Lapostolle [33] and Sacherer [34] showed that 2-dimensional KV- envelope equations are applicable for arbitrary beams with elliptical symmetry, while beam sizes and beam emittances are represented by corresponding rms values. Sacherer also demonstrated that 3D envelope equations are only weakly dependent on type of beam distribution, and can be used for non-uniformly charged beams with ellipsoidal symmetry [34]. The 3D equations for beam envelopes $R_x = \sqrt{5\langle x^2 \rangle}$, $R_y = \sqrt{5\langle y^2 \rangle}$ $R_z = \sqrt{5\langle \zeta^2 \rangle}$, Eqs. (B.18) - (B.20), can be written as

$$\frac{d^2 R_x}{dt^2} = -\frac{q\beta c}{m\gamma} G(z) R_x + f_x(R_x, R_y, R_z), \tag{2.1}$$

$$\frac{d^2 R_y}{dt^2} = \frac{q\beta c}{m\gamma} G(z) R_y + f_y(R_x, R_y, R_z), \tag{2.2}$$

$$\frac{d^2 R_z}{dt^2} = f_z(R_x, R_y, R_z), \tag{2.3}$$

where the functions $f_x, f_y, f_z$ are

$$f_x(R_x, R_y, R_z) = \frac{\varepsilon_x^2 c^2}{\gamma^2 R_x^3} + \frac{\Omega^2 R_x}{2} + 3\frac{I}{I_c} \frac{c^2 M_x \lambda}{\gamma^3 R_y R_z}, \tag{2.4}$$

$$f_y(R_x, R_y, R_z) = \frac{\varepsilon_y^2 c^2}{\gamma^2 R_y^3} + \frac{\Omega^2 R_y}{2} + 3\frac{I}{I_c} \frac{c^2 M_y \lambda}{\gamma^3 R_x R_z}, \tag{2.5}$$

$$f_z(R_x, R_y, R_z) = -\Omega^2 R_z + \frac{\varepsilon_z^2 c^2}{\gamma^6 R_z^3} + 3\frac{I}{I_c} \frac{M_z c^2 \lambda}{\gamma^3 R_x R_y}. \tag{2.6}$$

$m$ and $q$ are the mass and charge of particles, $\beta$ and $\gamma$ are beam velocity and energy, $G(z)$ is the gradient of quadrupole lenses, $\varepsilon_x = 5\varepsilon_{x\_rms}$, $\varepsilon_x = 5\varepsilon_{x\_rms}$, $\varepsilon_x = 5\varepsilon_{x\_rms}$ are normalized beam emittances, $\Omega$ is the longitudinal oscillation frequency, Eq. (B.4), $I$ is the beam current, $I_c = 4\pi\varepsilon_o mc^3 / q$ is the characteristic beam current, $\lambda$ is the RF wavelength, and $M_x$, $M_y$, $M_z$ are ellipsoid coefficients, Eqs. (B10) – (B.12). Below we consider a typical case where the longitudinal oscillation frequency is changing slowly with respect to the variation of alternative-sign gradient of quadrupole lenses. Importantly, envelope equations contain rapidly oscillating parts determined by the function $G(z)$, and weakly oscillating functions $f_x$, $f_y$, $f_z$ due to variation of beam sizes $R_x$, $R_y$, $R_z$. However, when phase advances of particle oscillations per focusing period do not exceed 60°, variation of beam sizes of the matched beam at the period of the structure are small, and functions $f_x$, $f_y$, $f_z$ can be approximately assumed to be $z$ -

independent. The gradient of the focusing field with focusing period $S$ can be expanded into Fourier series:

$$G(z) = \sum_{n=1}^{\infty} g_n \sin \frac{2\pi n z}{S}. \tag{2.7}$$

Using the expansion, Eq. (2.7), let us rewrite the transverse envelope equations as

$$\frac{d^2 R_x}{dt^2} = -\sum_{n=1}^{\infty} F_n R_x \sin(\omega_n t) + f_x(R_x, R_y, R_z), \tag{2.8}$$

$$\frac{d^2 R_y}{dt^2} = \sum_{n=1}^{\infty} F_n R_y \sin(\omega_n t) + f_y(R_x, R_y, R_z), \tag{2.9}$$

where the harmonics of expansion of the quadrupole field are

$$F_n = g_n \frac{q\beta c}{m\gamma}, \qquad \omega_n = \frac{2\pi n \beta c}{S}. \tag{2.10}$$

Within the framework of the averaging method [35], solutions of differential equations of second order, containing fast oscillating and slow variable terms, can be represented as

$$R_x(z) = \bar{R}_x(z) + \xi_x(z), \qquad R_y(z) = \bar{R}_y(z) + \xi_y(z), \tag{2.11}$$

where $\bar{R}_x(z)$, $\bar{R}_y(z)$ are slow variable functions with respect to fast oscillation frequency determined by alternative-sign gradients, and $\xi_x(z)$, $\xi_y(z)$ are small-amplitude rapidly oscillating functions. For the smooth part of the solution, application of the averaging method gives:

$$\frac{d^2 \bar{R}_x}{dt^2} = -\frac{1}{2} \sum_{n=1}^{\infty} \frac{F_n^2}{\omega_n^2} \bar{R}_x + f_x(\bar{R}_x, \bar{R}_y, R_z), \tag{2.12}$$

$$\frac{d^2 \bar{R}_y}{dt^2} = -\frac{1}{2} \sum_{n=1}^{\infty} \frac{F_n^2}{\omega_n^2} \bar{R}_y + f_y(\bar{R}_x, \bar{R}_y, R_z). \tag{2.13}$$

Equations (2.12), (2.13) can be re-written as

$$\frac{d^2 \bar{R}_x}{dt^2} - \frac{\varepsilon_x^2 c^2}{\gamma^2 \bar{R}_x^3} + (\Omega_r^2 - \frac{\Omega^2}{2}) \bar{R}_x - 3 \frac{I}{I_c} \frac{c^2 M_x \lambda}{\gamma^3 \bar{R}_y R_z} = 0, \tag{2.14}$$

$$\frac{d^2 \bar{R}_y}{dt^2} - \frac{\varepsilon_y^2 c^2}{\gamma^2 \bar{R}_y^3} + (\Omega_r^2 - \frac{\Omega^2}{2}) \bar{R}_y - 3 \frac{I}{I_c} \frac{c^2 M_y \lambda}{\gamma^3 \bar{R}_x R_z} = 0, \tag{2.15}$$

where the frequency of transverse linear oscillations

$$\Omega_r^2 = \frac{1}{2}(\frac{q\beta c}{m\gamma})^2 \sum_{n=1}^{\infty} \frac{g_n^2}{\omega_n^2}. \tag{2.16}$$

Small-amplitude fast components $\xi_x(z)$, $\xi_y(z)$ are determined by quickly oscillating terms only:

$$\xi_x = \sum_{n=1}^{\infty} \frac{F_n \sin\omega_n t}{\omega_n^2} \bar{R}_x, \qquad \xi_y = -\sum_{n=1}^{\infty} \frac{F_n \sin\omega_n t}{\omega_n^2} \bar{R}_y. \tag{2.17}$$

Taking only the first terms in Eqs. (2.17), these functions are

$$\xi_x = \upsilon_{max} \bar{R}_x \sin(\frac{2\pi\beta ct}{S}), \quad \xi_y = -\upsilon_{max} \bar{R}_y \sin(\frac{2\pi\beta ct}{S}), \quad \upsilon_{max} = \frac{q}{m\gamma \beta c} \frac{S^2 g_1}{4\pi^2}. \tag{2.18}$$

Accordingly, the solutions to the envelope equations in smooth approximation can be written as

$$R_x(z) = \bar{R}_x(z)[1 + \upsilon_{max} \sin(2\pi \frac{z}{S})], \tag{2.19}$$

$$R_y(z) = \bar{R}_y(z)[1 - \upsilon_{max} \sin(2\pi \frac{z}{S})]. \tag{2.20}$$

From equations (2.19), (2.20) the slopes of beam envelopes are

$$\frac{dR_x(z)}{dz} = 2\pi \upsilon_{max} \frac{\bar{R}_x}{S} \cos(2\pi \frac{z}{S}), \qquad \frac{dR_y(z)}{dz} = -2\pi \upsilon_{max} \frac{\bar{R}_y}{S} \cos(2\pi \frac{z}{S}). \tag{2.21}$$

Twiss parameters of the beam $\alpha_{x,y}$, $\beta_{x,y}$ are related to envelope parameters through well-known expressions

$$\beta_{x,y} = \frac{R_{x,y}^2}{\varepsilon_{x,y}}(\beta\gamma), \qquad \alpha_{x,y} = -\frac{dR_{x,y}}{dz}\sqrt{\frac{\beta_{x,y}}{\varepsilon_{x,y}}(\beta\gamma)}. \tag{2.22}$$

Under the adopted approximations, the longitudinal beam size, $R_z(z)$, is a slow changing variable with respect to transverse envelopes, $R_x(z)$, $R_y(z)$. To this end, the phase advance of transverse oscillations per focusing period $S$ is $\mu_o = \Omega_r S/(\beta c)$, or

$$\mu_o = \frac{qS}{m\gamma}\sqrt{\frac{1}{2}\sum_{n=1}^{\infty}\frac{g_n^2}{\omega_n^2}}. \tag{2.23}$$

Therefore, the envelope equations for smoothed 3D beam envelopes $\bar{R}_x$, $\bar{R}_y$, $R_z$ can be written as

$$\frac{d^2 \bar{R}_x}{dz^2} - \frac{\varepsilon_x^2}{(\beta\gamma)^2 \bar{R}_x^3} + \frac{\mu_s^2}{S^2} \bar{R}_x - 3\frac{I}{I_c} \frac{M_x \lambda}{\beta^2 \gamma^3 \bar{R}_y R_z} = 0, \qquad (2.24)$$

$$\frac{d^2 \bar{R}_y}{dz^2} - \frac{\varepsilon_y^2}{(\beta\gamma)^2 \bar{R}_y^3} + \frac{\mu_s^2}{S^2} \bar{R}_y - 3\frac{I}{I_c} \frac{M_y \lambda}{\beta^2 \gamma^3 \bar{R}_x R_z} = 0, \qquad (2.25)$$

$$\frac{d^2 R_z}{dz^2} - \frac{\varepsilon_z^2}{(\beta\gamma^3)^2 R_z^3} + \frac{\mu_{oz}^2}{S^2} R_z - 3\frac{I}{I_c} \frac{M_z \lambda}{\beta^2 \gamma^3 \bar{R}_x \bar{R}_y} = 0, \qquad (2.26)$$

where the phase advance of transverse oscillations of a synchronous particle at the period of focusing structure in presence of an RF field:

$$\mu_s = \mu_o \sqrt{1 - \frac{\mu_{oz}^2}{2\mu_o^2}}. \qquad (2.27)$$

In general, equations (2.24) - (2.26) describe envelope oscillations of an unmatched beam, with slow variation of average radius $\bar{R}_x(z)$, $\bar{R}_y(z)$, $R_z(z)$, superimposed onto fast oscillations, Eqs. (2.19), (2.20), with the period of the focusing structure. For a FODO focusing structure, the phase advance of transverse oscillations, $\mu_o$, and amplitude of fast oscillations, $\upsilon_{max}$, are [36]

$$\mu_o = \frac{S}{2D} \sqrt{1 - \frac{4}{3} \frac{D}{S} \frac{qG_o D^2}{m\gamma\beta c}}, \qquad \upsilon_{max} = \frac{2}{\pi^2 \sqrt{1 - \frac{4}{3} \frac{D}{S}}} \frac{\sin(\pi \frac{D}{S})}{(\pi \frac{D}{S})} \mu_o. \qquad (2.28)$$

Furthermore, in the thin lens approximation, $D \ll S$, Eq. (2.28) gives for the amplitude of small oscillations:

$$\upsilon_{max} \approx \frac{2}{\pi^2} \mu_o = 0.2026 \mu_o. \qquad (2.29)$$

Conversely, for $D = 0.5S$ (FD structure),

$$\upsilon_{max} = \frac{4\sqrt{3}}{\pi^3} \mu_o \approx 0.223 \mu_o. \qquad (2.30)$$

In the case of beam focusing by a periodic sequence of solenoids with period $S$, length $D$, and magnetic field $B_o$, the phase advance of transverse oscillations per focusing period and ripple amplitude are given by [37]

$$\mu_o = \theta \sqrt{\frac{S}{D}} \sqrt{1 - \frac{\theta^2}{6}[1 - \frac{1}{2}(\frac{D}{S} + \frac{S}{D})]}, \qquad (2.31)$$

$$\upsilon_{max} = \frac{\theta^2}{2\pi^3} (\frac{S}{D})^2 \sin(\pi \frac{D}{S}), \qquad (2.32)$$

$$\theta = \frac{qB_o D}{2mc\beta\gamma}. \tag{2.33}$$

Although the above discussion only considers the explicit cases of FODO and solenoid channels, it is worth noting that after averaging, all structures are well characterized by the values of the phase advance, $\mu_o$, and depths of variation of envelopes, $\upsilon_{\max}$.

### 3. Six-Dimensional Matched Beam

Equations (2.24) - (2.26) allow us to determine matched beam conditions, when transverse beam envelopes are periodic with the period of focusing structure. In many cases, transverse beam emittances are close to each other, which allows us to simplify the analysis of beam matching. Correspondingly, let us now consider matched beam with equal transverse emittances $\varepsilon_x = \varepsilon_y = \varepsilon$ and equal averaged transverse sizes $\overline{R}_x = \overline{R}_y = R$. In such a matched beam, transverse variations of envelopes are determined by fast oscillation components, Eqs. (2.18), while averaged components of envelopes are constant, $\overline{R}_x'' = \overline{R}_y'' = \overline{R}_z'' = 0$. Accordingly, a beam with equal transverse radii is a uniformly charged spheroid with ellipsoid coefficients, Eqs. (B.10) - (B.12)

$$M_x = M_y = \frac{(1-M_z)}{2}. \tag{3.1}$$

From Eq. (B.9), the potential of this uniformly charged spheroid is

$$U_b(r,\zeta) = -\frac{\rho}{2\varepsilon_o}[M_z \gamma^2 \zeta^2 + \frac{1-M_z}{2} r^2], \tag{3.2}$$

where the spheroid coefficient is

$$M_z = \frac{\gamma R^2 R_z}{2} \int_o^\infty \frac{ds}{(R^2+s)(\gamma^2 R_z^2 + s)^{3/2}}. \tag{3.3}$$

Evaluating the integral in Eq. (3.3) for a prolate spheroid with eccentricity $\varsigma = \sqrt{1-(R/\gamma R_z)^2}$ gives [38]:

$$M_z = \frac{1-\varsigma^2}{\varsigma^2}(\frac{1}{2\varsigma}\ln\frac{1+\varsigma}{1-\varsigma}-1). \tag{3.4}$$

Conversely, for an oblate spheroid with eccentricity $\varsigma = \sqrt{(R/\gamma R_z)^2 - 1}$, the parameter $M_z$ is [38]

$$M_z = \frac{1+\varsigma^2}{\varsigma^2}(1-\frac{\text{arctg}\,\varsigma}{\varsigma}). \tag{3.5}$$

Figure 1 illustrate the dependencies $M_z(R/\gamma R_z)$. It is further worth noting that for typical beam parameters, $0.2 < R/(\gamma R_z) < 1$, coefficient $M_z$ can be approximated as

$$M_z \approx \frac{R}{3\gamma R_z}. \tag{3.6}$$

With the above definitions in place, the equilibrium envelope equations are:

$$-\frac{\varepsilon^2}{(\beta\gamma)^2 R^3} + \frac{\mu_s^2}{S^2}R - \frac{3}{2}\frac{I}{I_c(\beta\gamma)^3 R}(\frac{\beta\lambda}{R_z})(1-M_z) = 0, \tag{3.7}$$

$$-\frac{\varepsilon_z^2}{(\beta\gamma^3)^2 R_z^3} + \frac{\mu_{oz}^2}{S^2}R_z - 3\frac{I}{I_c}\frac{\beta\lambda}{(\beta\gamma)^3 R^2}M_z = 0. \tag{3.8}$$

Equations (3.7), (3.8) connect equilibrium beam sizes $R$, $R_z$, beam emittances $\varepsilon$, $\varepsilon_z$, and beam current $I$, with accelerator parameters characterized by phase advances of transverse, $\mu_s$, and longitudinal, $\mu_{oz}$, oscillations. Importantly, equations (3.7), (3.8) can be rewritten concisely as

$$\varepsilon = \beta\gamma\frac{\mu_t R^2}{S}, \tag{3.9}$$

$$\varepsilon_z = \beta\gamma^3\frac{\mu_z R_z^2}{S}, \tag{3.10}$$

where depressed transverse, $\mu_t$, and depressed longitudinal, $\mu_z$, phase advances per focusing period are:

$$\mu_t^2 = \mu_s^2[1 - \frac{3}{2}\frac{I}{I_c(\beta\gamma)^3}(\frac{\beta\lambda}{R_z})(\frac{S}{R})^2\frac{(1-M_z)}{\mu_s^2}], \tag{3.11}$$

$$\mu_z^2 = \mu_{oz}^2[1 - \frac{3I}{I_c(\beta\gamma)^3}(\frac{\beta\lambda}{R_z})(\frac{S}{R})^2\frac{M_z}{\mu_{oz}^2}]. \tag{3.12}$$

The ratio of Eqs. (3.9) and (3.10) gives the relationship between beam parameters and depressed phase advances for equilibrium bunch:

$$\frac{\varepsilon}{\varepsilon_z} = \frac{\mu_t}{\mu_z}(\frac{R}{\gamma R_z})^2. \tag{3.13}$$

Equations (3.11), (3.12) can be re-written as

$$\mu_t^2 = \mu_s^2 - \frac{\rho}{\rho_c}(\frac{1-M_z}{2}), \tag{3.14}$$

$$\mu_z^2 = \mu_{oz}^2 - \frac{\rho}{\rho_c} M_z, \tag{3.15}$$

where $\rho$ is the space charge density of the ellipsoidal bunch, Eq. (B.7), and $\rho_c$ is the characteristic space charge density:

$$\rho_c = \frac{I_c \beta^2 \gamma^3}{4\pi c S^2}. \tag{3.16}$$

Cumulatively, equations (3.14), (3.15) indicate that the reduction of space charge effect in both transverse and longitudinal directions can be achieved with lowering the ratio $\rho/\rho_c$.

The equilibrium equation for transverse beam radius, Eq. (3.7), can be rewritten as

$$\left(\frac{R}{R_{ot}}\right)^4 - 2b_t \left(\frac{R}{R_{ot}}\right)^2 - 1 = 0, \tag{3.17}$$

where we have introduced an equilibrium radius of a beam with vanishing current $I = 0$

$$R_{ot} = \sqrt{\frac{\varepsilon S}{\beta \gamma \mu_s}}, \tag{3.18}$$

and transverse space charge parameter

$$b_t = \frac{3}{2} \frac{I}{I_c \beta \gamma} \left(\frac{R_{ot}}{\varepsilon}\right)^2 \left(\frac{\beta \lambda}{2R_z}\right)(1 - M_z). \tag{3.19}$$

Equation (3.17) is satisfied by

$$R = R_{ot} \sqrt{b_t + \sqrt{1 + b_t^2}}. \tag{3.20}$$

Accordingly, equation (3.20) can be understood as a solution for a matched beam radius of continuous beam [13], where the "effective" current of the beam is

$$I_{eff} = \frac{3}{2}(1 - M_z)\left(\frac{\beta \lambda}{2R_z}\right) I. \tag{3.21}$$

Taking into account that for most beams $M_z \approx 0.2 - 0.3$, the "effective" current of a typical beam is $I_{eff} \approx I/B$, where beam bunching factor is

$$B = \frac{2R_z}{\beta \lambda}. \tag{3.22}$$

The equilibrium equation for the longitudinal beam radius, Eq. (3.8), can be rewritten as

$$(\frac{R_z}{R_{oz}})^4 - b_z(\frac{R_z}{R_{oz}})^3 - 1 = 0, \tag{3.23}$$

where equilibrium longitudinal beam radius at negligible beam current is

$$R_{oz} = \sqrt{\frac{\varepsilon_z S}{\beta \gamma^3 \mu_{oz}}}, \tag{3.24}$$

and longitudinal space charge parameter

$$b_z = 3\gamma^3 M_z \frac{I}{I_c} \frac{\lambda R_{oz}^3}{R^2 \varepsilon_z^2}. \tag{3.25}$$

Together, equations (3.20), (3.23) determine matched beam sizes $R$, $R_z$ through given normalized beam emittances, $\varepsilon$, $\varepsilon_z$, beam current $I$, beam momentum $\beta\gamma$ in a linac with wavelength $\lambda$ and undepressed phase advances $\mu_o$, $\mu_{oz}$ per focusing period $S$. To this end, it is important to understand that the solution for the transverse equilibrium beam size, $R$, depends on longitudinal equilibrium beam size, $R_z$, and the solution for $R_z$ in turn depends on $R$. Thus, to find stationary matched beam conditions for a bunched beam in an RF field, equations (3.20), (3.23) have to be solved together. Practically speaking, the search for a solution should be performed until the sum of squares of equations (3.20), (3.23)

$$\chi = (\frac{R}{R_{ot}} - \sqrt{b_t + \sqrt{1 + b_t^2}})^2 + [(\frac{R_z}{R_{oz}})^4 - b_z(\frac{R_z}{R_{oz}})^3 - 1]^2 \tag{3.26}$$

reaches small value, typically $\chi \approx 10^{-7}$. Upon finding equilibrium beam sizes, $R$, $R_z$, equations (2.19) - (2.20), provide matched conditions for oscillating beam envelopes. Crucially, we note that the solutions to equations (3.20), (3.23) exist for any combination of beam and structure parameters as long as depressed phase advances, Eqs. (3.11), (3.12) are $\mu_t \geq 0$, $\mu_z \geq 0$.

Figure 2 illustrates matching of the beam with the accelerator when the beam is at the distance of $S/4$ from the middle of the first quadrupole. At this position, the beam has equal transverse sizes $R_x = R_y = R$ determined by Eq. (3.20), with slopes of beam envelopes defined as

$$\frac{dR_x}{dz} = 2\pi \upsilon_{max} \frac{R}{S}, \qquad \frac{dR_y}{dz} = -2\pi \upsilon_{max} \frac{R}{S}. \tag{3.27}$$

Deviation of beam envelopes from matched conditions results in un-periodic beam oscillations which manifests in an additional growth of phase space occupied by the beam.

## 4. Six-Dimensional Matched Beam with Equal Space Charge Depression

Among the infinitely large number of possible matched beam solutions, the system of Eqs. (3.7), (3.8) for equilibrium beam sizes has a specific solution, when space charge depression of particle oscillations are the same in transverse and longitudinal directions. Eliminating beam current from both equations (3.7), (3.8), they can be combined together to give

$$\frac{2\mu_s^2}{(1-M_z)}[1-\frac{\varepsilon^2}{(\beta\gamma\frac{\mu_s R^2}{S})^2}] = \frac{\mu_{oz}^2}{M_z}[1-\frac{\varepsilon_z^2}{(\beta\gamma^3\frac{\mu_{oz}R_z^2}{S})^2}]. \qquad (4.1)$$

A solution to Eq. (4.1) can be written as a combination of two equations

$$\frac{2\mu_s^2}{(1-M_z)} = \frac{\mu_{oz}^2}{M_z}, \qquad (4.2)$$

$$\frac{\varepsilon}{\varepsilon_z} = \frac{\mu_s}{\mu_{oz}}(\frac{R}{\gamma R_z})^2, \qquad (4.3)$$

where we independently equate terms inside and outside square brackets in Eq. (4.1). Equation (4.2) has a solution [39]

$$M_{ze} = \frac{\mu_{oz}^2}{2\mu_o^2}. \qquad (4.4)$$

The value of $M_{ze}$ depends on the ratio of bunch semi-axis, Eq. (3.3), and determines the shape of the equilibrium bunch. According to Eq. (4.4), this shape is quantified by longitudinal and transverse phase advances of accelerator structure. On the other hand, substitution of Eq. (4.4) into Eqs. (3.11), (3.12) reveals the condition for equal space-charge depression of transverse and longitudinal oscillation frequencies in both planes:

$$\eta^2 = \frac{\mu_t^2}{\mu_s^2} = \frac{\mu_z^2}{\mu_{oz}^2} = 1 - \frac{3}{2\mu_o^2}\frac{I}{I_c(\beta\gamma)^3}(\frac{\beta\lambda}{R_z})(\frac{S}{R})^2. \qquad (4.5)$$

Further substitution of the relation $\mu_t/\mu_z = \mu_s/\mu_{oz}$ into Eq. (3.13) results in Eq. (4.3). Therefore, the solution, Eq. (4.4), also satisfies Eq. (4.3). Beam radii satisfying Eqs. (4.2), (4.3), are determined by Eqs. (3.20), (3.23) taking into account the auxiliary condition given by Eq. (4.4):

$$(\frac{R}{R_{ot}})^4 - \tilde{b}_t(\frac{R}{R_{ot}}) - 1 = 0, \qquad \tilde{b}_t = \frac{3}{2}(\frac{\mu_s}{\mu_o})^2(\frac{\lambda R_{ot}}{\varepsilon^2})(\frac{R}{\gamma R_z}), \qquad (4.6)$$

$$(\frac{R_z}{R_{oz}})^4 - \tilde{b}_z(\frac{R_z}{R_{oz}}) - 1 = 0, \qquad \tilde{b}_z = \frac{3}{2}\gamma\frac{\mu_{oz}^2}{\mu_o^2}\frac{I}{I_c}(\frac{\lambda R_{oz}}{\varepsilon_z^3})(\frac{\gamma R_z}{R})^2. \qquad (4.7)$$

Notably, equations (4.6), (4.7) are nonlinear equations of fourth order, which determine the increase of equilibrium beam sizes with space charge. Figure 3 illustrates the associated dependences, Eqs. (4.6), (4.7).

Condition (4.4) expresses an equilibrium bunch with equal space charge suppression of oscillation frequencies in both transverse and longitudinal directions. While here we have arrived at this result as a limiting case of 3D envelope equations, the same expression was derived in Ref. [39] from different grounds – namely self-consistent beam analysis for intense bunched beam in an RF field. Equal space charge depression is accepted as a design criterion for the ESS linac [40-42].

Terms in square brackets in Eq. (4.1) according to Eqs. (3.9), (3.10) are

$$1 - \frac{\varepsilon^2}{(\beta\gamma \frac{\mu_s R^2}{S})^2} = 1 - \frac{\mu_t^2}{\mu_s^2}, \tag{4.8}$$

$$1 - \frac{\varepsilon_z^2}{(\beta\gamma^3 \frac{\mu_{oz} R_z^2}{S})^2} = 1 - \frac{\mu_z^2}{\mu_{oz}^2}. \tag{4.9}$$

For strong space charge depression, $\mu_t / \mu_s \ll 1$, $\mu_z / \mu_{oz} \ll 1$, expressions within square brackets in Eq. (4.1) are close to unity, and Eq. (4.4) is a unique condition for profile of an equilibrium beam, regardless of other parameters.

Combination of Eqs. (4.2), (4.3) provides a relationship between ratio of beam emittances and that of beam sizes for equal space charge depression:

$$\frac{\varepsilon}{\varepsilon_z} = (\frac{R}{\gamma R_z})^2 \sqrt{\frac{1 - M_{ze}}{2M_{ze}}}. \tag{4.10}$$

Taking into account the approximation for an ellipsoidal coefficient $M_z$ within the range $0.2 < R/(\gamma R_z) < 1$, Eq. (3.6), the Eq. (4.10) can be written as:

$$\frac{\varepsilon}{\varepsilon_z} \approx \sqrt{\frac{3}{2}} (\frac{R}{\gamma R_z})^{3/2} \sqrt{1 - \frac{R}{3\gamma R_z}}. \tag{4.11}$$

Alternatively, equation (4.11) can be re-written in terms of ellipsoidal coefficient $M_{ze}$

$$\frac{\varepsilon}{\varepsilon_z} \approx \frac{9}{\sqrt{2}} M_{ze}^{3/2} \sqrt{1 - M_{ze}}. \tag{4.12}$$

Figure 4 illustrates the relationship between beam radii and beam emittances for beams with equal space charge depression, Eq. (4.10), and that pertaining to the well-known equipartitioning condition [43]

$$\frac{\varepsilon}{\varepsilon_z} = \frac{R}{\gamma R_z}. \tag{4.13}$$

The aforementioned equipartitioning condition, Eq. (4.13), upon being combined with Eq. (3.13) can be also expressed as:

$$\frac{\varepsilon}{\varepsilon_z} = \frac{\mu_z}{\mu_t}. \tag{4.14}$$

We remark that the equipartitioning condition entails equal thermal spreads of particle momentum in transverse and longitudinal directions. As can be readily seen in the Figure 4, the equal space charge depression condition, Eq. (4.10), is different from the equipartitioning condition, Eq. (4.13), although two are close to one-another. We thus speculate that the equipartitioning condition serves as a proxy for the more rigorous condition of equal transverse and longitudinal space charge depression.

## 5. Self-Consistent Beam Dynamics

Analysis of beam matching presented in previous Sections was performed within a linear approximation of particle dynamics around a synchronous trajectory with the additional approximation of beam space charge field corresponding of that of a uniformly charged ellipsoid. In Ref. [39], the self-consistent numerical analysis of high-intensity bunched beam equilibrium was performed for the case where the beam size is limited by the separatrix size. A key aim of this Section is to relax that assumption and obtain a purely analytic description of a bunched beam with arbitrary size contained within the separatrix.

For the problem at hand, the previously employed linear approximation to particle dynamics is no longer valid and the generalized Hamiltonian of averaged particle motion is given by

$$H = \frac{p_x^2 + p_y^2}{2m\gamma} + \frac{p_\zeta^2}{2m\gamma^3} + qU_{ext} + q\frac{U_b}{\gamma^2}, \tag{5.1}$$

where the potential of external field is

$$U_{ext} = \frac{E}{k_z}[I_o(\frac{k_z r}{\gamma})\sin(\varphi_s - k_z\zeta) - \sin\varphi_s + k_z\zeta\cos\varphi_s] + \frac{m\gamma}{q}\Omega_r^2 \frac{r^2}{2}. \tag{5.2}$$

Hamiltonian, Eq. (5.1), with potential function, Eq. (5.2), is a result of averaging of the equations of motion such that the fast oscillating term, which is proportional to the gradient of the focusing field, $G(z)$, is replaced by a constant focusing term proportional to the square of transverse oscillation frequency, $\Omega_r^2$. Let us note, that in Eq. (5.2) the value of synchronous phase is negative, $\varphi_s < 0$. Moreover, for a matched stationary beam, the space charge potential is time-independent, and, therefore, the Hamiltonian is a constant of motion. The general approach to finding a stationary, self-consistent beam distribution function is to represent it as a function of the constant of motion, $f = f(H)$, and then to solve Poisson's equation for the unknown space charge potential of the beam [13]. In Ref. [39] the self-consistent problem for a high-brightness beam in an RF field was solved assuming 6D beam distribution function is an exponential function of the Hamiltonian, $f = f_o \exp(-H/H_0)$. In particular, it was shown, that, to the first approximation, the self-consistent space charge potential of an intense beam is opposite to the external potential:

$$U_b = -\frac{\gamma^2}{1+\delta}U_{ext}, \tag{5.3}$$

where $\delta$ is a small parameter

$$\delta \approx \frac{1}{b_\varphi k} \ll 1, \tag{5.4}$$

the value $k \approx 1.5 - 3$ is the bunch shape form-factor, $b_\varphi$ is the dimensionless brightness of the beam

$$b_\varphi = \frac{2}{\beta\gamma} \frac{I}{I_c B} (\frac{a}{\varepsilon})^2, \tag{5.5}$$

$B$ is the bunching factor, Eq. (3.22), and $a$ is the radius of beam pipe aperture. Comparison of Eq. (5.5) with Eq. (3.11) gives the following expression for the dimensionless beam brightness

$$b_\varphi = (\frac{R}{a})^2 [(\frac{\mu_s}{\mu_t})^2 - 1]. \tag{5.6}$$

For a common ratio $R/a \approx 0.5$ adopted in derivations of Ref. [39], one can write

$$b_\varphi k = (\frac{\mu_s}{\mu_t})^2 - 1. \tag{5.7}$$

The space charge potential of self-consistent bunch is then

$$U_b = -\gamma^2 [1 - (\frac{\mu_t}{\mu_s})^2] U_{ext}. \tag{5.8}$$

Equation (5.8) indicates that the space charge field of a stationary bunch compensates for the external field in the beam interior. This phenomenon is well-known within the context of non-neutral plasma physics as Debye shielding. With these developments, the Hamiltonian of a self-consistent beam takes the form

$$H = \frac{p_x^2 + p_y^2}{2m\gamma} + \frac{p_\zeta^2}{2m\gamma^3} + q(\frac{\mu_t}{\mu_s})^2 U_{ext}. \tag{5.9}$$

Equation (5.9) indicates that in presence of an intense bunched beam, the stationary longitudinal phase space of the beam becomes narrow in momentum spread, while the phase width of the distribution remains, to leading order, unchanged. This is in qualitative agreement with the study of Ref. [13], where the self-consistent problem for a stationary bunched beam was solved numerically for the longitudinal phase space particle distribution. In particular, the resulting bunch in this work was approximated by a hard-edged cylinder, where the particle density was constant in every cross-section of the cylinder, but depended on the longitudinal position. As such, numerical results from Ref. [13] indicated that in the space charge-dominated regime, the separatrix of the longitudinal phase space was substantially reduced in momentum, while remaining almost unchanged in phase width, which is fully consistent with the picture entailed by Eq. (5.9).

The self-consistent space charge density distribution of a matched beam can be derived from Poisson's equation:

$$\rho(r,\zeta) = -\varepsilon_o \left[\frac{1}{r}\frac{\partial}{\partial r}(r\frac{\partial U_b}{\partial r}) + \frac{\partial^2 U_b}{\gamma^2 \partial \zeta^2}\right]. \tag{5.10}$$

Substitution of Eq. (5.8) into Eq. (5.10) gives the constant value of space charge density within the bunch:

$$\rho(r,\zeta) = 2\gamma^2 \frac{m\gamma}{q}\Omega_r^2 [1-(\frac{\mu_t}{\mu_s})^2]. \tag{5.11}$$

Equation (5.9) indicates that transverse and longitudinal oscillation frequencies are suppressed by space charge in the same proportion. In fact, referring back to the Eqs. (B.1), (B.2) for small paraxial bunch, it is trivial to see that the external potential, Eq. (5.2), is given by

$$U_{ext}(r,\zeta) = \frac{m\gamma^3}{q}\Omega^2 \frac{\zeta^2}{2} + \frac{m\gamma}{q}\Omega_r^2(1-\frac{\Omega^2}{2\Omega_r^2})\frac{r^2}{2}. \tag{5.12}$$

Substitution of Eq. (5.12) into Eq. (5.8) gives for the potential of a small self-consistent stationary bunch:

$$U_b = -\frac{\rho}{2\varepsilon_o}\left[\frac{\Omega^2}{2\Omega_r^2}(\gamma\zeta)^2 + (1-\frac{\Omega^2}{2\Omega_r^2})\frac{r^2}{2}\right]. \tag{5.13}$$

Notably, this potential is that of uniformly charged spheroid, Eq. (3.2), where the coefficient of the spheroid

$$M_{ze} = \frac{\Omega^2}{2\Omega_r^2} = \frac{\mu_{oz}^2}{2\mu_o^2}. \tag{5.14}$$

corresponds to a beam with equal space charge depression (see Eq. (4.4)). Therefore, the self-consistent solution for a stationary bunch results in a bunch where the space charge potential is opposite to the external potential, which in turn results in equal space charge depression in transverse and longitudinal directions.

In a more general case, the shape of self-consistent bunch can be different from a spheroid. Nevertheless, from Eq. (5.8) it follows, that, the space charge potential of the beam in the limit of strong space charge is the same function of coordinates, as the external potential, Eq. (5.2), with the opposite sign. Therefore, the equation $U_{ext}(r,\zeta) = const$ gives a family of equipotential lines of the space charge field of the bunch:

$$I_o(\frac{k_z r}{\gamma})\sin(\varphi_s - k_z\zeta) - \sin\varphi_s + k_z\zeta\cos\varphi_s + \frac{m\gamma}{q}\frac{k_z}{E}\Omega_r^2\frac{r^2}{2} = const, \tag{5.15}$$

where $k_z\zeta = \psi$ is the phase deviation from a synchronous particle. Furthermore, in the general case, the bunch boundary is not an equipotential surface; therefore Eq. (5.15) does not coincide with the bunch profile. To find the self-consistent bunch profile, consider a uniformly populated bunch with boundary $r(\zeta)$, defined by the following nonlinear equation

$$I_o(\frac{k_z r}{\gamma})\sin(\varphi_s - k_z \zeta) - \sin\varphi_s + k_z \zeta \cos\varphi_s + C(k_z r)^2 = const. \quad (5.16)$$

In Ref. [39] it was shown that a uniformly charged bunch with boundary, Eq. (5.16), creates a space charge field, which is approximately opposite to the external RF field acting at the bunch, as is required by self-consistent solution for space charge potential, Eq. (5.8). The value of the constant in right hand of Eq. (5.16) can be determined from the condition, that at the axis, $r=0$, the maximum deviation from the synchronous particle in positive direction is $\zeta_{max} = R_z$:

$$const = \sin(\varphi_s - k_z R_z) - \sin\varphi_s + k_z R_z \cos\varphi_s. \quad (5.17)$$

Similarly, the value of $C$ is determined from the condition that at $\zeta = 0$, the beam radius is $r = R$:

$$C = \frac{1}{(k_z R)^2}[\sin(\varphi_s - k_z R_z) + k_z R_z \cos\varphi_s - I_o(\frac{k_z R}{\gamma})\sin\varphi_s]. \quad (5.18)$$

Finally, after substitution of Eqs. (5.17), (5.18) into Eq. (5.16), the shape of the bunch $(r,\zeta)$ is determined by the nonlinear equation

$$I_o(\frac{k_z r}{\gamma})\sin(\varphi_s - k_z \zeta) + k_z \zeta \cos\varphi_s + C(k_z r)^2 - \sin(\varphi_s - k_z R_z) - k_z R_z \cos\varphi_s = 0. \quad (5.19)$$

In Ref. [39], the similar equation was obtained for a beam occupying the whole separatrix, while coefficient $C$ was determined numerically. Note that for a short paraxial bunch, $k_z R_z \ll 1$, $k_z R \ll 1$, the value of constant $C$, Eq. (5.18), is

$$C = \frac{|\sin\varphi_s|}{2}[(\frac{R_z}{R})^2 + \frac{1}{2\gamma^2}], \quad (5.20)$$

and the shape of the bunch, Eq. (5.19) is transformed into ellipsoid:

$$(\frac{\zeta}{R_z})^2 + (\frac{r}{R})^2 = 1. \quad (5.21)$$

Figure 5 illustrates the profile of a matched beam, Eq. (5.19), for various ratios of $k_z R_z / |\varphi_s|$, for $\varphi_s$ = -30°. Clearly, the equilibrium bunch shape in real space resembles the shape of the separatrix in longitudinal phase space. Accordingly, the bunch shape is transformed from an ellipsoidal shape for $k_z R_z / |\varphi_s| \ll 1$ into a "separatrix" shape as $k_z R_z / |\varphi_s|$ approaching unity.

Figure 6 illustrates dynamics of a beam with $k_z R_z / |\varphi_s| = 0.8$ with initial ellipsoidal and matched, Eq. (5.19), beam profile in a structure with $\mu_{oz} = 29.8°$, $\mu_z = 16.4°$, $\mu_o = 76.7°$, $\mu_t = 35.4°$. These simulations were performed using particle-in-cell code BEAMPATH [44]. As can be seen, beam with an initial ellipsoidal shape tends to change its profile to progressively resemble that, described by Eq. (5.19), while the shape of the matched beam is approximately conserved. Importantly, redistribution of beam density results in growth of 6D phase space volume. Figure 7 illustrates the relative growth of the 6D phase space volume, $V_6 = \sqrt{\det \sigma}$, which is calculated through determinant of 6D beam sigma-matrix:

$$\vec{\sigma} = \begin{pmatrix} <x^2> & <xp_x> & <xy> & <xp_y> & <x\delta\tau> & <x\delta\gamma> \\ <p_x x> & <p_x^2> & <p_x y> & <p_x p_y> & <p_x \delta\tau> & <p_x \delta\gamma> \\ <yx> & <yp_x> & <y^2> & <yp_y> & <y\delta\tau> & <y\delta\gamma> \\ <p_y x> & <p_y p_x> & <p_y y> & <p_y^2> & <p_y \delta\tau> & <p_y \delta\gamma> \\ <\delta\tau x> & <\delta\tau p_x> & <\delta\tau y> & <\delta\tau p_y> & <(\delta\tau)^2> & <\delta\tau \delta\gamma> \\ <\delta\gamma x> & <\delta\gamma p_x> & <\delta\gamma y> & <\delta\gamma p_y> & <\delta\gamma \delta\tau> & <(\delta\gamma)^2> \end{pmatrix}, \quad (5.22)$$

where $\delta\tau = (t - t_s)c/\lambda$, and $\delta\gamma = \gamma - \gamma_s$. As can be immediately understood, 6D phase space volume growth is smaller in case of matched beam profile, Eq. (5.19). Therefore, Eq. (5.19) provides additional specification on matched beam conditions together with solution of Eqs. (3.20), (3.23).

## 6. Computer Simulations of Matched Beams

Analyses presented in previous sections, provide conditions for a matched beam in 6D phase space. An important advantage of the developed analytical framework is that it allows for a meaningful comparison between various beam equilibria, and their effects on beam phase-space growth. To this end, a particularly interesting line of inquiry is to determine the role of equipartitioning with respect to other beam equilibria. The optimal combination of accelerator and beam parameters resulting in minimization of phase space growth and reduction of beam losses remains a key problem in high-intensity beam physics. Equipartitioning has been considered as the only beam equilibrium which can minimize emittance growth [43, 45]. The latest experimental data [46], however, indicate that equipartitioning is not necessarily required to minimize beam losses.

In order to compare the effect of 6D phase space distortion in matched beams with various ratios of beam radii and beam emittances, we performed a series of computer simulations for three different dependencies of beam sizes and emittances:

- equal space charge depression, Eq. (4.10),
- equipartitioning, Eq. (4.13),
- equal emittances, $\varepsilon = \varepsilon_z$.

Simulations were performed using the BEAMPATH code [44], keeping values of transverse, $\mu_o$, and longitudinal, $\mu_{oz}$, phase advances per focusing periods constant along the accelerating structure, with an initial Gaussian distribution in 6D phase space. Longitudinal beam size, $R_z$, and longitudinal emittance, $\varepsilon_z$, were selected to be the same for all simulations because the shape of the matched bunch changes with variation of $R_z$, as discussed in Section 5. Longitudinal undepressed phase advance in all cases was selected to be $\mu_{oz} = 29.8^{o}$, synchronous phase $\varphi_s = -30^{o}$, longitudinal bunch size 22º, and longitudinal space charge depression $\mu_z / \mu_{oz} = 0.2$. Within the chosen set of parameters, the beam current was varied only with transverse beam radius:

$$I = \frac{I_c}{3}(\beta\gamma)^3 (\frac{R_z}{\beta\lambda})(\frac{R}{S})^2 \frac{\mu_{oz}^2}{M_z}[1 - \frac{\mu_z^2}{\mu_{oz}^2}] \approx I_c (\beta\gamma^2)^2 \mu_{oz}^2 (\frac{R_z}{S})^2 (\frac{R}{\lambda})[1 - \frac{\mu_z^2}{\mu_{oz}^2}]. \tag{6.1}$$

Selected parameters gave us an opportunity to provide analysis within the linear part of sinusoidal RF field and compare simulation and analytical results closely. To this end, References [47, 48] suggest that strong space depression ~ 0.2 corresponds to area where resonant islands in stability charts disappear, and emittance growth due to "free energy" effect is dominant.

For the equal space charge depression study, simulations were performed with a constant value of space charge depression in both transverse and longitudinal directions:

$$\eta = \frac{\mu_t}{\mu_s} = \frac{\mu_z}{\mu_{oz}} = 0.2. \tag{6.2}$$

The ratio of the transverse beam size to the longitudinal beam size was varied within $0.4 < R/\gamma R_z < 1.4$ range. Every value of the ratio $R/\gamma R_z$ determined the ellipsoid coefficient, Eq. (3.3), which, in turn, defines the ratio of beam emittances according to Eq. (4.10). The value of transverse phase advance is then determined from Eq. (4.4) as

$$\mu_o = \frac{\mu_{oz}}{\sqrt{2M_{ze}}} \tag{6.3}$$

defined the quadrupole focusing structure. For the study of beam dynamics under the equipartitioning condition, each value of $R/\gamma R_z$ determined the transverse beam emittance, $\varepsilon = \varepsilon_z R/(\gamma R_z)$, as well as the depressed transverse phase advance $\mu_t = \mu_z (\varepsilon_z / \varepsilon)$ according to equipartitioning equations, Eqs. (4.13), (4.14). Transverse undepressed phase advance in presence of an RF field is determined from Eqs. (3.14), (3.15), (2.27):

$$\mu_s = \sqrt{\mu_t^2 + (\mu_{oz}^2 - \mu_z^2)(\frac{1-M_z}{2M_z})}, \qquad \mu_o = \sqrt{\mu_s^2 + \frac{\mu_{oz}^2}{2}}. \tag{6.4}$$

In case of equal emittances, $\varepsilon = \varepsilon_z$, each value of $R/\gamma R_z$ determined transverse depressed phase advance in presence of an RF field from Eq. (3.13):

$$\mu_t = \mu_z (\frac{\gamma R_z}{R})^2. \tag{6.5}$$

After that, the value of undepressed phase advance of focusing structure was determined from Eq. (6.4).

Figure 8 illustrates relative growth of 6D phase space volume in all cases. As can be seen, 6D phase space volume growth is correlated with the variation of transverse space charge depression

(see Fig. 9), while longitudinal space charge depression was kept constant $\mu_z/\mu_{oz} = 0.2$. Phase space growth was approximately constant in case of equal space-charge depression, but varied along equipartitioning and equal emittances conditions following variation of ratio $\mu_t/\mu_s$.

It is important to understand that the phase space volume growth is a result of beam phase density redistribution. To this end, the "free energy" model gives the following prescription for beam emittance growth of a spherical bunch with equal space charge depression [49]

$$\frac{\varepsilon}{\varepsilon_i} = \sqrt{1 + \frac{1}{3}(\frac{1}{\eta^2} - 1)\Delta U} = 1.75, \qquad (6.6)$$

where $\Delta U = 0.26$ is the normalized value of electrostatic bunch field variation for the beam with an initial Gaussian distribution, and $\eta = 0.2$ is the space charge depression factor, Eq. (6.2). Observable values of emittance growth in simulations for spherical bunch were $\varepsilon_x/\varepsilon_{xo} = 2.03$, $\varepsilon_y/\varepsilon_{yo} = 2.13$, $\varepsilon_z/\varepsilon_{zo} = 1.78$. The difference can be explained by a presence of tails in the resulting 6D phase space distribution. We note that the smallest value of 6D phase space volume growth was observed for the case of equal emittances, when, under selected parameters, the transverse space charge depression had the smallest value of $\mu_t/\mu_s \approx 0.58$.

Figures 10 - 12 illustrate evolution of equipartitioning parameter

$$T_e = (\frac{\varepsilon_z}{\varepsilon} \frac{R}{\gamma R_z})^2 \qquad (6.7)$$

and growth of 6D phase space volume versus focusing period of the structure for all modes of simulation. Clearly, there is a tendency for space charge dominated beam to evolve towards the equipartitioning condition $T_e = 1$ rapidly at the time of quick phase space volume growth associated with phase space density redistribution. After that, there are oscillations of the value of $T_e$ around individual asymptotic values. Nevertheless, beams with initial conditions $0.4 \le R/\gamma R_z \le 1.0$ tend to oscillate around equiapartitioning value $T_e = 1$ in a long-term, even when not being initially equipartitioned (see Figs 10, 12). Conversely, beams with initial conditions $R/\gamma R_z > 1.0$ tend to deviate from the value $T_e = 1$ in all cases, even when beam was initial equipartitioned (see Fig. 11). Thus, the performed simulations indicate that equipartitioning condition is not necessarily associated with smallest beam phase space growth. Instead, minimization of phase space volume growth can be achieved with optimization of space charge depression in transverse and longitudinal directions

## 7. Experimental Study of Beam Matching

The Los Alamos Neutron Science Center (LANSCE) accelerator facility (formerly known as LAMPF) started routine operation in 1972 as a 0.8 MW average proton beam power facility for meson physics research, and delivered high-power beam for a quarter century. Currently, the accelerator sends beams to five experimental areas (see Fig. 13 and Table 1). The accelerator is equipped with two independent injectors for H$^+$ and H$^-$ beams, merging at the entrance of a 201.25 MHz Drift Tube Linac (DTL). The DTL performs acceleration up to the energy of 100 MeV. After the DTL, the Transition Region beamline directs a 100 MeV proton beam to the Isotope Production Facility, while the H$^-$ beam is accelerated up to the final energy of 800 MeV in an 805-MHz Coupled Cavity Linac. The H⁻ beams, created by different time structure of a low-energy chopper, are distributed in the Switch Yard (SY) to four experimental areas: Lujan Neutron Scattering Center equipped with Proton Storage Ring (PSR), Weapon Neutron Research Facility (WNR), Proton Radiography Facility (pRad), and Ultra-Cold Neutron Facility (UCN).

Beam losses in the LANSCE accelerator are mainly determined by the most powerful 80 kW H⁻ beam, injected into Proton Storage Ring. Beam losses are controlled by various types of loss monitors. The main control is provided by Activation Protection (AP) detectors, which are one-pint size cans with a photomultiplier tube immersed in scintillator fluid. AP detectors integrate the signals and shut off the beam if the beam losses around an AP device exceed 100 nA of average current. The same devices are used as beam loss monitors (LM), where the signal is not integrated and therefore one can see a real-time of beam loss across the beam pulse.

Detailed study of beam dynamics and determination of optimal conditions for 6D beam matching to LANSCE DTL linac were done numerically using TRACE-3D and PARMILA codes multiple times within long-term history of accelerator operation [17, 50 - 52]. Good agreement between beam simulation and LANSCE beam measurement were found for accelerated rms beam emittances (within 25% - 50%), while simulated beam losses were overestimated with respect to observations by an order of magnitude. Beam matching to accelerator is performed at the entrance of Drift Tube Linac at the beam energy of 750 keV. Transverse matching of the beam is achieved through beam emittance measurements and subsequent transformation of beam ellipses into that determined by DTL lattice using four matching quadrupoles. Longitudinal matching is performed through adjustments of amplitudes and phases of 2 - cavity buncher system to maximize beam capture into DTL.

The Drift Tube Linac consists of 4 tanks with output energies of 5 MeV, 41 MeV, 73 MeV, and 100 MeV, respectively. Originally designed for operation with a synchronous phase of -26°, the linac was historically retuned for -32°, -23°, -22°, -32° tank synchronous phases with field amplitudes of 98%, 96%, 94%, and 98% of nominal values to minimize beam spill. Both H⁻ and proton beams are captured with efficiencies of 75% - 82% into the Drift Tube Linac, so initially 20% - 25% of the beam is lost in the beginning of Tank 1. Subsequent beam losses of 0.1% - 1% in the DTL are determined by additional uncaptured particles and by expansion of the phase-space volume occupied by the beam. Coupled Cavity Linac includes 44 accelerating modules (modules 5-48). Modules 5-12 contain 4 accelerating tanks each, while modules 13 - 48 contain 2 tanks each. Typical averaged integrated beam losses along CCL linear accelerator from 100 MeV to 800 MeV are 2 x 10⁻³.

Entrance DTL lattice is a FODO structure with quadrupoles of the length $D$ = 3.522 cm and field gradient $G \approx 7600$ kGs / cm occupying each drift tube between accelerating gaps with accelerating gradient $E = E_o T$ =1.3 MV/m. Drift Tube Linac is an Alvarez - type with focusing period $S = 2\beta\lambda$. Using these numbers, the values of phase advances at the entrance of DTL according to Equations (2.27), (2.28), (B.6) are $\mu_o = 55°$, $\mu_{oz} = 42.8°$, $\mu_s = 45.9°$, and the value of relative envelope variation is $\upsilon_{max} = 0.222$. The first quadrupole in the DTL, $Q_o$, has an irregular value of length $D_{Qo} = 0.705\, D$ and gradient $G_{Qo} = 0.61\, G$. Numerical integration of periodic solution for beam envelopes in DTL structure to the entrance of $Q_o$ quadrupole gives the following numbers for Twiss parameters of the zero-current matched beam in front of DTL:

$$\beta_x = 26.01 \text{ cm}, \qquad \beta_y = 5.95 \text{ cm}, \qquad \alpha_x = 0.0271, \qquad \alpha_y = -0.0575. \tag{7.1}$$

In presence of space charge, the values of matched conditions depend on combination of beam emittance and beam current. LANSCE H⁻ beam has a typical transverse normalized rms beam emittance $\varepsilon_{x\_rms} \approx \varepsilon_{y\_rms} = 0.02\, \pi$ cm mrad, while estimated longitudinal rms beam emittance after

2-cavity bunching system is $\varepsilon_{z\_rms} = 0.15\,\pi$ cm mrad. Average matched zero-current beam sizes according to Eqs. (3.18), (3.24) are $R_{ot} = 0.193$ cm, $R_{oz} = 0.546$ cm. The present-day operation value of H⁻ beam current is $I = 10$ mA, while future plans include increase of beam current by a factor of 1.5. Calculations of matched conditions for bunched beam with current $I = 15$ mA through simultaneous solution of Eqs. (3.20), (3.23) gives for space charge parameters, Eqs. (3.19), (3.25) $b_t = 0.304$, $b_z = 0.167$, and for matched beam sizes

$$R = 0.223 \text{ cm}, \qquad R_z = 0.566 \text{ cm}. \tag{7.2}$$

The relative increases of matched beam sizes due to space charge are therefore $R/R_{ot} = 1.15$, $R_z/R_{oz} = 1.03$. The values of space charge tune depressions are

$$\frac{\mu_t}{\mu_s} = 0.74, \qquad \frac{\mu_z}{\mu_{oz}} = 0.91, \tag{7.3}$$

and equipartitioning parameter $T_e = 2.9$. Recalculations of matched beam to the beginning of entrance DTL quadrupole give the following Twiss parameters for matched beam including space charge forces

$$\beta_x = 37.83 \text{ cm}, \quad \beta_y = 8.46 \text{ cm}, \quad \alpha_x = 0.617, \quad \alpha_y = -0.186. \tag{7.4}$$

Numerical simulations [52] show that longitudinal beam size $R_z$ at the entrance of DTL is a factor of 1.5 larger than that required by beam space charge matching conditions, Eq. (7.2), therefore, beam remains mismatched along linac. Insufficient longitudinal matching is due to utilization of simple 2-cavity bunching system currently adopted in linac. Improvement of beam matching is expected with substitution of existing bunching system with RFQ - based Front End [53].

Historically, the zero-current Twiss parameters, Eq. (7.1), are used for beam matching in LANSCE accelerator as a compromise solution for injection of various beams. Such approach provides certain mismatch at the entrance of the machine for each specific beam. In order to quantify the effect of mismatch, the experimental study was done to observe variation of H⁻ beam losses along accelerator as a function of transverse beam mismatch at the entrance to Drift Tube Linac from the values, Eq. (7.4). The mismatch parameter is $F = 0.5(F_x + F_y)$, where

$$F_x = \sqrt{\frac{1}{2}(P_x + \sqrt{P_x^2 - 4}) - 1}, \tag{7.5}$$

and $P_x = \beta_m \gamma_b + \beta_b \gamma_m - 2\alpha_m \alpha_b$ is the parameter indicating overlap of $x$ - beam ellipses with matched beam Twiss parameters $\alpha_m$, $\beta_m$, $\gamma_m$, and that of the beam, $\alpha_b$, $\beta_b$, $\gamma_b$, and similarly for $P_y$. Beam ellipses were varied at the entrance of linac using matching quadrupoles, providing change of mismatch parameter within wide range of values $0 \leq F \leq 3$, while beam losses were measured along accelerator. Figure 14 illustrates results of measurements. Whereas significant increase of beam losses was observed at the level of mismatch factor $F \approx 1$, the single beam trip might happen at smaller value of mismatch factor $F = 0.6$. Mismatch between zero-current beam matching values, Eq. (7.1), and non-zero current values, Eq. (7.4) is $F = 0.27$. Relative insensitivity of beam losses to initial beam mismatch can be explained by the fact that normalized transverse acceptance of the LANSCE linac is significantly larger than rms beam emittance (by a factor of ~ 50) along the machine.

## 8. Summary

Analytical conditions for a matched beam in 6D phase space were derived from 3D envelope equations and from self-consistent analysis of an intense beam in RF field. It is demonstrated that a matched bunch shape is transformed from an ellipsoidal profile to a "separatrix"- type shape in real space when longitudinal beam size becomes comparable with separatrix size. Simulation results confirm the predictions of the developed theoretical model. A new class of beam equilibrium with equal space charge depression in transverse and longitudinal directions is studied. Correspondingly, a numerical comparison was undertaken for three different types of equilibria (equipartitioning, equal space-charge depression, and equal emittances). It is shown that minimization of six-dimensional phase space volume growth is not related to any specific beam equilibrium, like equipartitioning, but is instead related to minimization of space charge depression. Results of the performed analysis are applied to experimental study of beam matching at LANSCE linac, providing quantitative threshold for initial beam mismatch to avoid significant beam losses in linac.

## Acknowledgments

Author is indebted to Konstantin Batygin for useful discussions and help in preparation of manuscript. Work is supported by US DOE under contract 89233218CNA000001.

## Appendix A. 6D Hamiltonian of Particle Motion in RF field

Consider single particle motion in an RF accelerator, in presence of space charge forces. The Hamiltonian associated with particle motion is

$$K = c\sqrt{m^2c^2 + (P_x - qA_x)^2 + (P_y - qA_y)^2 + (P_z - qA_z)^2} + qU_a + qU_{el} + qU_b, \qquad (A.1)$$

where $m$ and $q$ are the mass and charge of particles, $U_a$ is the potential of accelerating field, $U_{el}$ is the potential of electrostatic focusing lenses, $U_b$ is the scalar potential of field of the beam, $P_{x,y,z}$ are components of particle canonical momentum, and $A_{x,y,z}$ are the components of the vector potential of focusing and space charge fields of the beam. We describe accelerator field as an equivalent traveling wave with an effective potential

$$U_a = \frac{E}{k_z} I_o\left(\frac{k_z r}{\gamma}\right) \sin(\omega t - k_z z), \qquad (A.2)$$

where $E = E_o T$ is the amplitude of the equivalent traveling wave, $I_o(k_z r / \gamma)$ is the modified Bessel function of zero-th order, $E_o$ is the average field at accelerating gap, $T$ is the transit time factor, $k_z = 2\pi/(\beta\lambda)$ is the wave number, and $\omega = 2\pi c/\lambda$ is the circular RF frequency. For further analysis, let us introduce new variables

$$p_\zeta = P_z - P_s, \qquad \zeta = z - z_s, \qquad (A.3)$$

which define deviation away from a synchronous particle with position $z_s$ and longitudinal momentum $P_s$. The generating function of the transformation is

$$F(\zeta, P_z, t) = -(\zeta + z_s)(P_z - P_s), \qquad (A.4)$$

and the relationships between old and new variables are determined by canonical transformations

$$p_\zeta = -\frac{\partial F}{\partial \zeta}, \qquad z = -\frac{\partial F}{\partial P_z}. \qquad (A.5)$$

In terms of these new variables, the new Hamiltonian is given by

$$T = c\sqrt{m^2c^2 + (P_x - qA_x)^2 + (P_y - qA_y)^2 + (P_s + p_\zeta - qA_z)^2} + qU_a + qU_{el} + qU_b + \frac{\partial F}{\partial t}. \qquad (A.6)$$

Consider separately the expression for the square root in Hamiltonian:

$$s(p_x, p_y, p_\eta) = \sqrt{(mc)^2 + p_x^2 + p_y^2 + (P_s + p_\eta)^2}, \qquad (A.7)$$

where the components of the canonical momentum are substituted by that of the mechanical momentum, $p_x = P_x - qA_x$, $p_y = P_y - qA_y$, and an additional variable $p_\eta = p_\zeta - qA_z$ is introduced. Longitudinal momentum of the synchronous particle is typically much larger than transverse particle momentum and longitudinal momentum spread, i.e. $P_s \gg p_x$, $P_s \gg p_y$, $P_s \gg p_\eta$. Accordingly, let us expand the expression for $s(p_x, p_y, p_\eta)$ up to the order of $p_x^2$, $p_y^2$, $p_\eta^2$:

$$s = \sqrt{(mc)^2 + P_s^2} + \frac{\partial s}{\partial p_x} p_x + \frac{\partial s}{\partial p_y} p_y + \frac{\partial s}{\partial p_\eta} p_\eta + \frac{1}{2} \frac{\partial^2 s}{\partial p_x^2} p_x^2$$

$$+ \frac{1}{2} \frac{\partial^2 s}{\partial p_y^2} p_y^2 + \frac{1}{2} \frac{\partial^2 s}{\partial p_\eta^2} p_\eta^2 + \frac{1}{2} \frac{\partial^2 s}{\partial p_x \partial p_y} p_x p_y + \frac{1}{2} \frac{\partial^2 s}{\partial p_x \partial p_\eta} p_x p_\eta + \frac{1}{2} \frac{\partial^2 s}{\partial p_y \partial p_\eta} p_y p_\eta, \quad \text{(A.8)}$$

where all derivatives are taken at $p_x = 0$, $p_y = 0$, $p_\eta = 0$. Calculation of expansion, Eq. (A.7), gives:

$$c\sqrt{(mc)^2 + p_x^2 + p_y^2 + (P_s + p_\zeta)^2} = mc^2 \gamma + \frac{p_x^2}{2m\gamma} + \frac{p_y^2}{2m\gamma} + \frac{P_s p_\eta}{m\gamma} + \frac{p_\eta^2}{2m\gamma^3}, \quad \text{(A.9)}$$

where $\gamma$ is the energy of a synchronous particle:

$$\gamma = \sqrt{1 + \left(\frac{P_s}{mc}\right)^2}. \quad \text{(A.10)}$$

The time derivative of the generating function, Eq. (A.4), is:

$$\frac{\partial F}{\partial t} = \zeta \dot{P}_s - \dot{z}_s P_z + \dot{z}_s P_s + z_s \dot{P}_s, \quad \text{(A.11)}$$

where dot denotes a derivative over time. Taking into account that the velocity of a synchronous particle is $\dot{z}_s = P_s / (m\gamma)$, the following expressions in time derivative, Eq. (A.11), are:

$$\dot{z}_s P_z = \frac{P_s}{m\gamma}(P_s + p_\zeta), \qquad \dot{z}_s P_s = \frac{P_s^2}{m\gamma}, \quad \text{(A.12)}$$

and the time derivative of the generating function is therefore

$$\frac{\partial F}{\partial t} = \zeta \dot{P}_s - \frac{P_s p_\zeta}{m\gamma} + z_s \dot{P}_s. \quad \text{(A.13)}$$

Substitution of expansions, Eqs. (A.9), (A.11), into Eq. (A.6) gives for the new Hamiltonian, $H = T - mc^2 \gamma$:

$$H = \frac{(P_x - qA_x)^2}{2m\gamma} + \frac{(P_y - qA_y)^2}{2m\gamma} + \frac{(p_\zeta - qA_z)^2}{2m\gamma^3} + qU_a + qU_{el} + qU_b - \frac{qP_s A_z}{m\gamma} + \dot{P}_s(z_s + \zeta). \quad \text{(A.14)}$$

The term $\dot{P}_s z_s$ can be excluded, because it does not depend on canonical variables and, therefore, does not contribute to the equations of particle motion. The acceleration of a synchronous particle is determined by expression $\dot{P}_s = qE\cos\varphi_s$, where $\varphi_s$ is the synchronous phase. The term $\dot{P}_s \zeta$ can be combined with the accelerating potential:

$$qU_a + \dot{P}_s\zeta = \frac{qE}{k_z}[I_o(\frac{k_z r}{\gamma})\sin(\varphi_s - k_z\zeta) + k_z\zeta\cos\varphi_s]. \tag{A.15}$$

Let us now consider the following terms in the Hamiltonian, Eq. (A.14):

$$\frac{(p_\zeta - qA_z)^2}{2m\gamma^3} - \frac{qP_s A_z}{m\gamma} = \frac{p_\zeta^2}{2m\gamma^3} - \frac{qP_s A_z}{m\gamma}(1 + \frac{p_\zeta}{P_s\gamma^2} - \frac{qA_z}{2P_s\gamma^2}). \tag{A.16}$$

Because $p_\zeta \ll P_s$, $qA_z \ll P_s$, the second and the third terms in parentheses in Eq. (A.16) can be omitted:

$$\frac{qP_s A_z}{m\gamma}(1 + \frac{p_\zeta}{P_s\gamma^2} - \frac{qA_z}{2P_s\gamma^2}) \approx \frac{qP_s A_z}{m\gamma} = q\beta_s cA_z. \tag{A.17}$$

Longitudinal component of vector potential $A_z = A_{z\,magn} + A_{z\_beam}$ is a combination of that of magnetic lenses, $A_{z\,magn}$, and that of the beam, $A_{z\_beam}$. In a moving coordinate system, where particles are static, the vector potential is zero. According to Lorentz transformation, the longitudinal component of beam vector - potential in laboratory system is $A_{z\_beam} = U_b\beta_s/c$. Therefore, under the adopted assumptions, the Hamiltonian, Eq. (A.14) becomes:

$$H = \frac{(P_x - qA_x)^2}{2m\gamma} + \frac{(P_y - qA_y)^2}{2m\gamma} + \frac{p_\zeta^2}{2m\gamma^3} + q(U_{el} - \beta cA_{z\,magn}) +$$

$$\frac{qE}{k_z}[I_o(\frac{k_z r}{\gamma})\sin(\varphi_s - k_z\zeta) + k_z\zeta\cos\varphi_s] + q\frac{U_b}{\gamma^2}. \tag{A.18}$$

Consider separately structures with quadrupole focusing and with longitudinal magnetic focusing. In the absence of longitudinal magnetic field, transverse components of the vector potential are $A_x = 0$, $A_y = 0$. Therefore, the transverse components of canonical momentum coincide with that of mechanical momentum: $p_x = P_x$, $p_y = P_y$. The term $U_{el} - \beta cA_{z\,magn}$ is the potential of a focusing structure of multipole lenses of an arbitrary order. We remark that, for quadrupole focusing,

$$U_{el} - \beta cA_{z\,magn} = \beta cG(z)\frac{x^2 - y^2}{2}, \tag{A.19}$$

where $G(z)$ is the gradient of focusing structure. However, the electrostatic quadrupole is equivalent to the magnetic one if it's gradient is $G_{el} = \beta c G$. The Hamiltonian for particle motion in RF field with quadrupole focusing is therefore

$$H = \frac{p_x^2 + p_y^2}{2m\gamma} + \frac{p_\zeta^2}{2m\gamma^3} + \frac{qE}{k_z}[I_o(\frac{k_z r}{\gamma})\sin(\varphi_s - k_z\zeta) + k_z\zeta\cos\varphi_s] + q\beta c G(z)\frac{x^2 - y^2}{2} + q\frac{U_b}{\gamma^2}. \quad (A.20)$$

In presence of a longitudinal magnetic field, the Hamiltonian, Eq. (A.18), is

$$H = \frac{(P_x - qA_x)^2}{2m\gamma} + \frac{(P_y - qA_y)^2}{2m\gamma} + \frac{p_\zeta^2}{2m\gamma^3} + \frac{qE}{k_z}[I_o(\frac{k_z r}{\gamma})\sin(\varphi_s - k_z\zeta) + k_z\zeta\cos\varphi_s] + q\frac{U_b}{\gamma^2}, \quad (A.21)$$

where the transverse components of the vector-potential are

$$A_x = -B\frac{y}{2}, \qquad A_y = B\frac{x}{2}, \quad (A.22)$$

and the components of the canonical momentum are

$$P_x = p_x - qB\frac{y}{2}, \qquad P_y = p_y + qB\frac{x}{2}. \quad (A.23)$$

Using transformation to the Larmor frame:

$$\hat{x} = x\cos\phi - y\sin\phi, \quad (A.24)$$

$$\hat{y} = x\sin\phi + y\cos\phi, \quad (A.25)$$

$$\hat{P}_x = P_x\cos\phi - P_y\sin\phi, \quad (A.26)$$

$$\hat{P}_y = P_y\cos\phi + P_x\sin\phi, \quad (A.27)$$

the Hamiltonian corresponding to particle motion in a magnetic field and an RF field is

$$H = \frac{\hat{P}_x^2 + \hat{P}_y^2}{2m\gamma} + \frac{p_\zeta^2}{2m\gamma^3} + \frac{qE}{k_z}[I_o(\frac{k_z r}{\gamma})\sin(\varphi_s - k_z\zeta) + k_z\zeta\cos\varphi_s] + m\gamma\omega_L^2\frac{r^2}{2} + q\frac{U_b}{\gamma^2}, \quad (A.28)$$

where the Larmor frequency is

$$\omega_L(z) = \frac{qB_z(z)}{2m\gamma}. \quad (A.29)$$

Hamiltonian, Eq. (A.29), is similar to Eq. (A.20) with the exception that it is defined in the Larmor frame, while, the Hamiltonian for a quadrupole channel, Eq. (A.20) is determined in the laboratory frame.

# Appendix B. Dynamics Around Synchronous Particle

Consider the dynamics of particles around a synchronous particle. Utilizing expansions

$$sin(\varphi_s - k_z\zeta) \approx sin\varphi_s - (k_z\zeta)cos\varphi_s - \frac{1}{2}(k_z\zeta)^2 sin\varphi_s, \tag{B.1}$$

$$I_o(\frac{k_z r}{\gamma}) \approx 1 + \frac{1}{4}(\frac{k_z r}{\gamma})^2, \tag{B.2}$$

the Hamiltonian, Eq. (A.20), becomes

$$H = \frac{p_x^2 + p_y^2}{2m\gamma} + \frac{p_\zeta^2}{2m\gamma^3} + m\gamma^3\Omega^2\frac{\zeta^2}{2} + q\beta cG(z)\frac{x^2-y^2}{2} - m\gamma\Omega^2\frac{(x^2+y^2)}{4} + q\frac{U_b}{\gamma^2}, \tag{B.3}$$

where the frequency of small amplitude linear oscillations around a synchronous particle is

$$\Omega = \sqrt{\frac{qEk_z|sin\varphi_s|}{m\gamma^3}}. \tag{B.4}$$

The dimensionless frequency of small amplitude oscillations is

$$\frac{\Omega}{\omega} = \sqrt{\frac{qE\lambda}{mc^2}\frac{|sin\varphi_s|}{2\pi\beta\gamma^3}}. \tag{B.5}$$

From Eq. (B.4), the phase advance of longitudinal oscillations per focusing period $S$, $\mu_{oz} = \Omega S/(\beta c)$, is given by

$$\mu_{oz} = \sqrt{2\pi(\frac{qE\lambda}{mc^2})\frac{|sin\varphi_s|}{\beta\gamma^3}}(\frac{S}{\beta\lambda}). \tag{B.6}$$

While there is no complete 6D self-consistent treatment of bunched beam dynamics in a linear field, it is common to formally include potential of a uniformly charged ellipsoid in equations of motion. We note, however, that such an ellipsoid is an exact self-consistent solution for a beam only in case of negligible beam phase space volume [54]. The space-charge density of the ellipsoid with semi-axes $R_x, R_y, R_z$ is given by

$$\rho = \frac{3}{4\pi}\frac{I\lambda}{cR_xR_yR_z}, \qquad \frac{x^2}{R_x^2} + \frac{y^2}{R_y^2} + \frac{\zeta^2}{R_z^2} \leq 1, \tag{B.7}$$

where $I\lambda/c = Q$ is the beam charge per bunch. In the moving system of coordinates $(\tilde{x}, \tilde{y}, \hat{\zeta})$, the vector potential of the beam is zero, $\tilde{\vec{A}}_b = 0$, and the beam field is described by a scalar potential [55]:

$$\tilde{U}_b(\tilde{x},\tilde{y},\tilde{\zeta}) = -\frac{\tilde{\rho}}{4\varepsilon_o} \tilde{R}_x \tilde{R}_y \tilde{R}_z \int_0^\infty \frac{(\frac{\tilde{x}^2}{\tilde{R}_x^2+s}+\frac{\tilde{y}^2}{\tilde{R}_y^2+s}+\frac{\tilde{\zeta}^2}{\tilde{R}_z^2+s})}{\sqrt{(\tilde{R}_x^2+s)(\tilde{R}_y^2+s)(\tilde{R}_z^2+s)}} ds, \qquad (B.8)$$

where $\tilde{x}=x$, $\tilde{y}=y$, $\tilde{\zeta}=\gamma\zeta$ are coordinates, $\tilde{R}_x = R_x$, $\tilde{R}_y = R_y$, $\tilde{R}_z = \gamma R_z$ are semi-axes of ellipsoid, and $\tilde{\rho} = \rho/\gamma$ is the space charge density of ellipsoid in moving system. According to the Lorentz transformation, the scalar potential in the laboratory system is $U_b = \gamma \tilde{U}_b$, which can be written as

$$U_b(x,y,\zeta) = -\frac{\rho}{2\varepsilon_o}[M_x x^2 + M_y y^2 + M_z \gamma^2 \zeta^2], \qquad (B.9)$$

where coefficients $M_x, M_y, M_z$ depend on the bunch shape:

$$M_x = \frac{1}{2}\int_0^\infty \frac{R_x R_y \gamma R_z\, ds}{(R_x^2+s)\sqrt{(R_x^2+s)(R_y^2+s)(\gamma^2 R_z^2+s)}}, \qquad (B.10)$$

$$M_y = \frac{1}{2}\int_0^\infty \frac{R_x R_y \gamma R_z\, ds}{(R_y^2+s)\sqrt{(R_x^2+s)(R_y^2+s)(\gamma^2 R_z^2+s)}}, \qquad (B.11)$$

$$M_z = \frac{1}{2}\int_0^\infty \frac{R_x R_y \gamma R_z\, ds}{(\gamma^2 R_z^2+s)\sqrt{(R_x^2+s)(R_y^2+s)(\gamma^2 R_z^2+s)}}. \qquad (B.12)$$

With the introduced space charge potential, Eq. (B.9), the Hamiltonian, Eq. (A.20), becomes

$$H = \frac{p_x^2+p_y^2}{2m\gamma} + \frac{p_\zeta^2}{2m\gamma^3} + m\gamma^3\Omega^2\frac{\zeta^2}{2} + q\beta cG(z)\frac{x^2-y^2}{2}$$

$$-m\gamma\Omega^2\frac{(x^2+y^2)}{4} - \frac{3mc^2 I\lambda}{2\gamma^2 I_c R_x R_y R_z}(M_x x^2 + M_y y^2 + M_z \gamma^2 \zeta^2), \qquad (B.13)$$

where $I_c = 4\pi\varepsilon_o mc^3/q = (A/Z)\cdot 3.13\cdot 10^7$ [A] is the characteristic current. Correspondingly, the equations of single particle motion derived from Hamiltonian, Eq. (B.13), are:

$$\frac{d^2 x}{dz^2} + [k_{x\psi}(z) - 3\frac{I}{I_c}\frac{\lambda M_x}{\beta^2\gamma^3 R_x R_y R_z}]x = 0, \qquad (B.14)$$

$$\frac{d^2 y}{dz^2} + [k_{y\psi}(z) - 3\frac{I}{I_c}\frac{\lambda M_y}{\beta^2\gamma^3 R_x R_y R_z}]y = 0, \tag{B.15}$$

$$\frac{d^2 \zeta}{dz^2} + [(\frac{\Omega}{\beta c})^2 - 3\frac{I}{I_c}\frac{\lambda M_z}{\beta^2\gamma^3 R_x R_y R_z}]\zeta = 0, \tag{B.16}$$

where $k_{x\psi}(z)$, $k_{y\psi}(z)$ are the focusing functions of a quadrupole focusing channel in presence of an RF field:

$$k_{x\psi}(z) = \frac{qG(z)}{mc\beta\gamma} - \frac{1}{2}(\frac{\Omega}{\beta c})^2, \qquad k_{y\psi}(z) = -\frac{qG(z)}{mc\beta\gamma} - \frac{1}{2}(\frac{\Omega}{\beta c})^2 \tag{B.17}$$

Equations of motion result in 3D envelope equations for a bunched beam:

$$\frac{d^2 R_x}{dz^2} - \frac{\varepsilon_x^2}{(\beta\gamma)^2 R_x^3} + k_{x\psi}(z) R_x - 3\frac{I}{I_c}\frac{M_x \lambda}{\beta^2\gamma^3 R_y R_z} = 0, \tag{B.18}$$

$$\frac{d^2 R_y}{dz^2} - \frac{\varepsilon_y^2}{(\beta\gamma)^2 R_y^3} + k_{y\psi}(z) R_y - 3\frac{I}{I_c}\frac{M_y \lambda}{\beta^2\gamma^3 R_x R_z} = 0, \tag{B.19}$$

$$\frac{d^2 R_z}{dz^2} - \frac{\varepsilon_z^2}{(\beta\gamma^3)^2 R_z^3} + \frac{\Omega^2(z)}{(\beta c)^2} R_z - 3\frac{I}{I_c}\frac{M_z \lambda}{\beta^2\gamma^3 R_x R_y} = 0, \tag{B.20}$$

where $\varepsilon_{x,y,z}$ are beam normalized emittances. For general distributions in 6D phase, the beam sizes and emittances are to be substituted by their equivalent values related to their rms values as $R_x = \sqrt{5\langle x^2 \rangle}$, $R_y = \sqrt{5\langle y^2 \rangle}$, $R_z = \sqrt{5\langle \zeta^2 \rangle}$, $\varepsilon_x = 5\varepsilon_{x\_rms}$, $\varepsilon_y = 5\varepsilon_{y\_rms}$, $\varepsilon_z = 5\varepsilon_{z\_rms}$ [32, 56].

# References

1. K. Hasegawa, "J-PARC Commissioning Results", Proceedings of 2005 Particle Accelerator Conference, Knoxville, Tennessee, p.220 (2005).

2. S. Henderson et al, "The Spallation Neutron Source Beam Commissioning and Initial Operations", ORNL Report ORNL/TM-2015/321 (2015).

3. A.M. Lombardi et al, "Linac4: from Initial Design to Final Commissioning", Proceedings of IPAC2017, Copenhagen, Denmark, TUYA1, p.1217 (2017).

4. P.N Ostroumov, "Physics Design of the 8 GeV H-Minus Linac", New Journal of Physics, 8. 281 (2006).

5. J.-F. Ostiguy at al, "Overview of Beam Optics in Project -X SC CW Linac", Proceedings of HB2010, Morschach, Switzerland, TUo1B04 (2010).

6. D. Vandeplassche et al, Proceedings of IPAC2011, 2718 (2011).

7. Zh. Li et al, "Physics Design of an Accelerator for an Accelerator-Driven Subcritical System", Phys Rev. Special Topics – Accelerators and Beams, 16 080101 (2013)

8. J. Knaster et al, Nucl. Fusion 53, 116001 (2013).

9. J. Wei, "The Very High Intensity Future", Proceedings of IPAC2014, Dresden, Germany, MOYBA01 , p.17 (2014).

10. M. Ball et al, "The PIP-II Conceptual Design Report", Fermilab Report FERMILAB-TM-2649-AD-APC (2017).

11. R. Garoby et al, "The European Spallation Source Design", Physica Scripta, 93 014001 (2018).

12. C. Plostinar et al, "Future High-Power Proton Drivers for Neutrino Beams", Proceedings of the IPAC2019, Melbourne, Australia, MOPRB045, p. 662 (2019).

13. I.M. Kapchinskiy, "Theory of Resonance Linear Accelerators", Harwood, 1983.

14. M. Reiser, "Theory and Design of Charged Particle Beams", Second Edition, Wiley-VCH Verlag GmbH (2008).

15. R.D. Ryne, "Finding Matching RMS Envelopes in RF Linacs: A Hamiltonian Approach", LA-UR-95-391, arXiv:acc-phys/9502001.

16. A.I. Balabin, G.N. Kropachev, "6D High Current Beam Matching at RFQ Entrance", Proceedings of EPAC 1994, p.1180 (1994).

17. F.E. Merrill, L.J. Rybarcyk, "Transverse Match of High Peak-Current Beam Into the LANSCE DTL Using Parmila", Proceedings of 1996 Linac Conference, (1996).

18. M. Pabst, K. Bongardt, L. Letchford, "Progress on Intense Proton Beam Dynamics and Halo Formation", Proceedings of EPAC 98, p 146-13.

19. N. Pichoff, "Envelope Modes of a Mismatched Bunched Beam", Note DAPNIA/SEA 98/44 (1998).

20. Ji Qiang, "Three-Dimensional Envelope Instability in Periodic Focusing Channels", PRAB 21, 034201 (2018).

21. R. S. Mills, K. R. Crandall, and J. A. Farrell, "Designing Self-Matching Linacs," Proc. 1984 Linac Conference, May 7-11, 1984, Darmstadt Report GSI-84-11, p.111 (1984).

22. M. Otani, et al., "Longitudinal Measurements and Beam Tuning in the J-PARC Linac MEBT1", Proceedings of the IPAC2019, Melbourne, Australia, MOPTS048, p. 968, (2019).

23. Ch. K. Allen, J.D. Galambos, W. Blokland, "Beam Profile Measurements and Matching at SNS: Practical Considerations and Accommodations", Proceedings of LINAC 2010, Tsukuba, Japan, TUP087, p.611 (2010).

24. M. Ikegami, "Beam Commissioning and Operation of the J-PARC Linac", Prog. Theor. Exp. Physics, 02B002 (2012).

25. K.R. Crandall, D.P. Rusthoi, "TRACE 3D Documentation", Los Alamos National Laboratory Report LA-UR -97-886 (1997).

26. http://irfu.cea.fr/dacm/logiciels/index.php.

27. K.L. Brown, F. Rothacker, D.C. Carey, Ch. Iselin, "TRANSPORT A Computer Program for Designing Charged Particle Beam Transport Systems", SLAC-91, Rev. 3, UC-28 (1983).

28. MAD - Methodical Accelerator Design, http://mad.web.cern.ch/mad.

29. F.J. Sacherer, "Transverse Space Charge Effects in Circular Accelerators" Ph.D. Thesis, University of California, Lawrence Radiation Laboratory Berkley, California, UCRL-18454, 1966.

30. P.M. Lapostolle, "Proton Linear Accelerators: A Theoretical and Historical Introduction", LA-11601-MS (1989).

31. P.M. Lapostolle, "Space Charge and High Intensity Effects in Radiofrequency Linacs", GANIL A.84-01 (1984)

32. T.P. Wangler, "RF Linear Accelerators", second ed., Wiley-VCH, Weinheim, 2005.

33. P.M. Lapostolle, "Possible Emittance Increase Through Filamentation due to Space Charge in Continuous Beams", Proceedings of the 1971 Particle Accelerator Conference, IEEE Transaction Nucl. Science NS-18, 1101 (1971).

34. F.J. Sacherer, "RMS Envelope Equations with Space Charge", Proceedings of the 1971 Particle Accelerator Conference, IEEE Transaction Nucl. Science NS-18, 1105 (1971).

35. L.D. Landau, E.M. Lifshitz, Mechanics, Elsevier, Amsterdam, 1974.

36. Y.K. Batygin, A. Scheinker, S. Kurennoy, Ch. Li "Suppression of space charge induced beam halo in nonlinear focusing channel", Nuclear Instruments and Methods in Physics Research A 816 (2016) 78–84.

37. Y.K. Batygin, "Dynamics of intense particle beam in axial-symmetric magnetic field", Nuclear Instruments and Methods in Physics Research A 772 (2015) 93–10A.

38. I.M. Kapchinsky, "Selected Topics in Ion Linac Theory", LA-UR-93-4192 (1993).

39. Y.K. Batygin "Self-consistent particle distribution of a bunched beam in RF field", Nuclear Instruments and Methods in Physics Research A 483 (2002) 611–626.

40. M. Eshraqi, J-M. Lagniel, "On the Choice of Linac Parameters for Minimal Beam Losses", Proceedings of IPAC 2013, Shanghai, China, p.1787-1789 (2013).

41. M.Eshraqi, "Cost Optimized Design of High Power Linacs", Proceedings of LINAC2014, Geneva, Switzerland, THIOA01, p. 785 (2014).

42. M. Eshraqi, J-M. Lagniel, "Choice of Linac parameters to minimize the space-charge effects", JINST, 15 P07024 (2020).

43. R. Jameson, IEEE Trans. Nucl. Sci. NS-28 (1981), 2406.

44. Y.K. Batygin, "Particle-in-cell code BEAMPATH for beam dynamics simulations in linear accelerators and beamlines", Nuclear Instruments and Methods in Physics Research A 539 (2005) 455–489.

45. R. Jameson, "An Approach to Fundamental Study of Beam Loss Minimization", Space Charge Dominated Beam Physics for Heavy Ion Fusion, AIP Conference Proceedings 480, Editor Y.K. Batygin, Saitama, Japan, p.21 (1998).

46. Y. Liu et al, "Progress of J-PARC Linac Commissioning", Proceedings of the IPAC2019, Melbourne, Australia, TUPIS027, p. 1990, (2019).

47. I. Hofmann, J. Qiang and R. D. Ryne, "Collective Resonance Model of Energy Exchange in 3D Nonequipartitioned Beams", Physical Review Letters, 86, 11, p. 2313 (2001).

48. I. Hofmann, G. Franchetti, O. Boine-Frankenheim, J.Qiang, R.D. Ryne, "Space Charge Resonance in Two and Three Dimensional Anisotropic Beams", Phys. Rev. Accelerators and Beams, 6, 024202 (2003).

49. I. Hofmann, J. Struckmeier, "Generalized Three-Dimensional Equations for the Emittance and Field Energy of High_Current Beams in Periodic Focusing Structures", Particle Accelerators, 1987, Vol. 21, pp. 69-96.

50. R. W. Garnett, R. S. Mills, T. P. Wangler, "Beam Dynamics Simulation of the LAMPF Linear Accelerator:, Proceedings of the Linear Accelerator Conference 1990, Albuquerque, New Mexico, USA, p.347 (1990).

51. R. W. Garnett, E. R. Gray, L. J. Rybarcyk, T. P. Wangler, "Simulation Studies of LAMPF Proton Linac", Proceedings of the 1995 Particle Accelerator Conference, p. 3185 (1995).

52. T. P. Wangler, F. Merrill, L. Rybarcyk, R.Ryne, "Space Charge in Proton Linacs", CP448, Workshop on Space Charge Physics in High Intensity Hadron Rings, Edited by A.U. Luccio and W.T. Weng, The American Insitute of Physics, p.3 (1998).

53. R.W. Garnett et al, "Status of the LANSCE Front End Upgrade", Proceedings of PAC2013, Pasadena, CA USA, MOPMA14, p.327 (2013).

54. Y.K. Batygin, "Self-Consistent Analysis of Three-Dimensional Uniformly Charged Ellipsoid with Zero Emittance", Physics of Plasmas, Vol. 6, Number 6, 3103- 3106 (2001).

55. W.D. MacMillan, "The Theory of Potential: Theoretical Mechanics", Dower, New York, 1956.

56. A. P.M. Lapostolle, CERN Report AR/Int SG/65-15, 1965.

Table 1: Beam Parameters of LANSCE Accelerator.

| Area | Rep. Rate (Hz) | Pulse Length (µs) | Current / bunch (mA) | Average current (µA) | Average power (kW) |
|---|---|---|---|---|---|
| Lujan | 20 | 625 | 10 | 100 | 80 |
| IPF | 100 | 625 | 4 | 230 | 23 |
| WNR | 100 | 625 | 25 | 2.5 | B.6 |
| pRad | 1 | 625 | 10 | <1 | <1 |
| UCN | 20 | 625 | 10 | 10 | 8 |

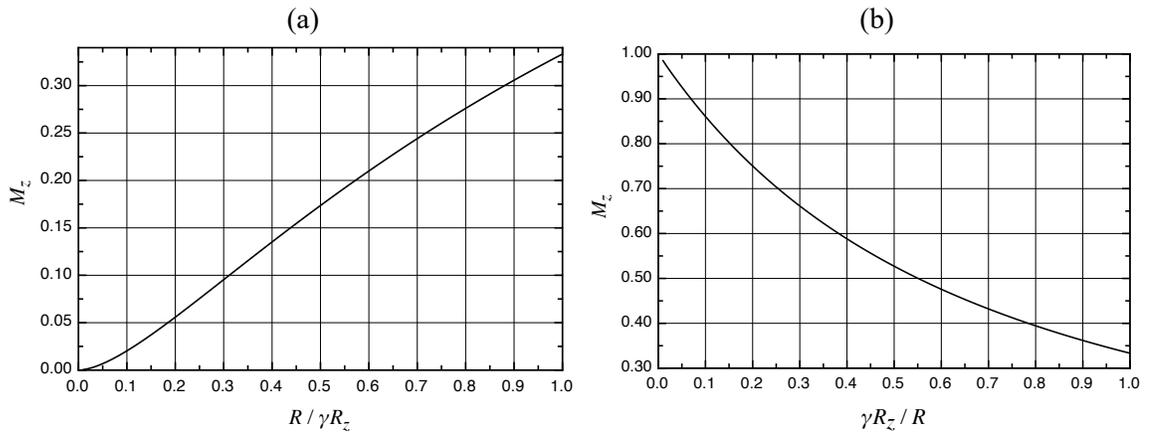

Figure 1 : Coefficient $M_z$ in spheroid potential, Eq. (3.3): (a) $R \leq \gamma R_z$, (b) $\gamma R_z \leq R$.

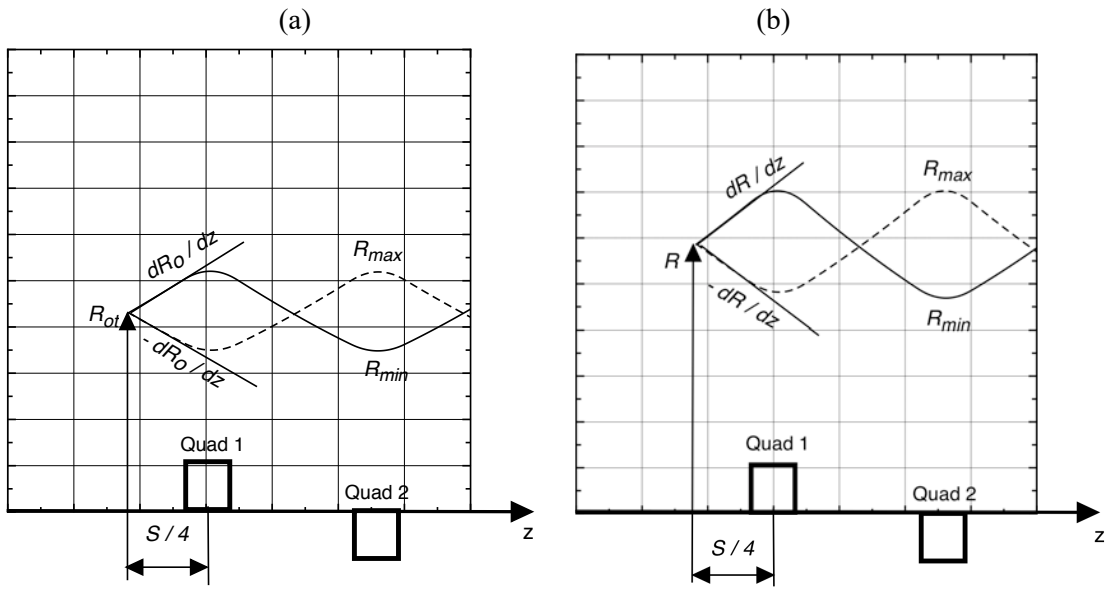

Figure 2: Transverse matching of the beam: (a) with negligible current, (b) with high-current.

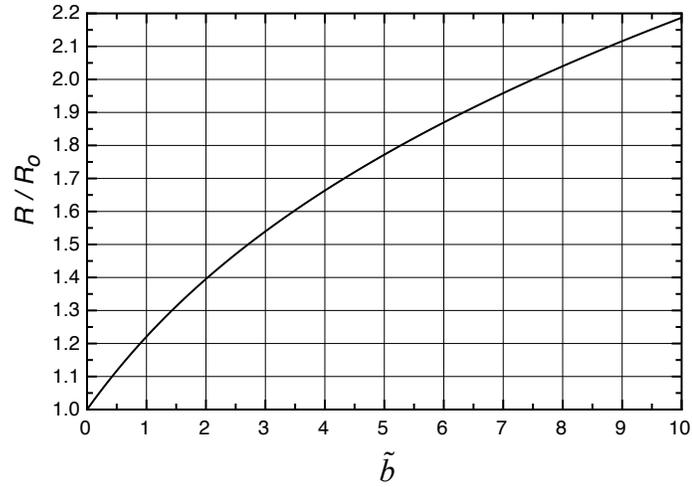

Figure 3: Increase of equilibrium beam size versus space charge parameter $\tilde{b}$ ($\tilde{b}_t$ or $\tilde{b}_t$ in Eqs. (4.6), (4.7)) for equal space charge depression mode.

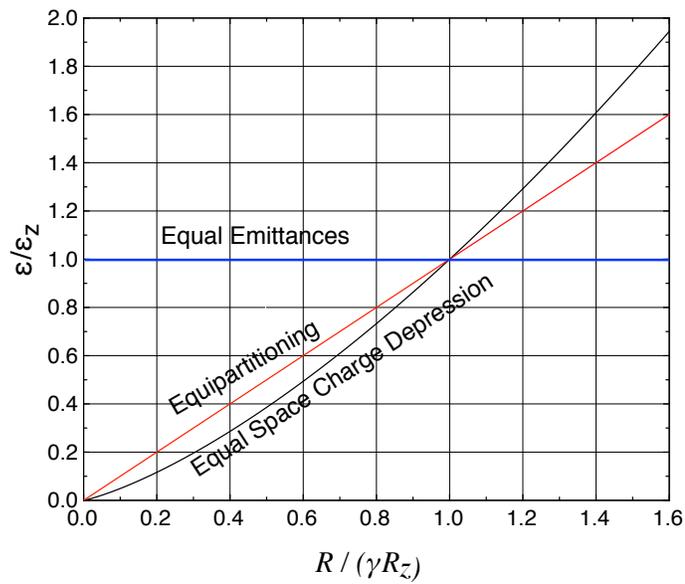

Figure 4: Ratio of beam emittances versus ratio of beam sizes for equipartitioning, equal space charge depression, and equal emittances modes.

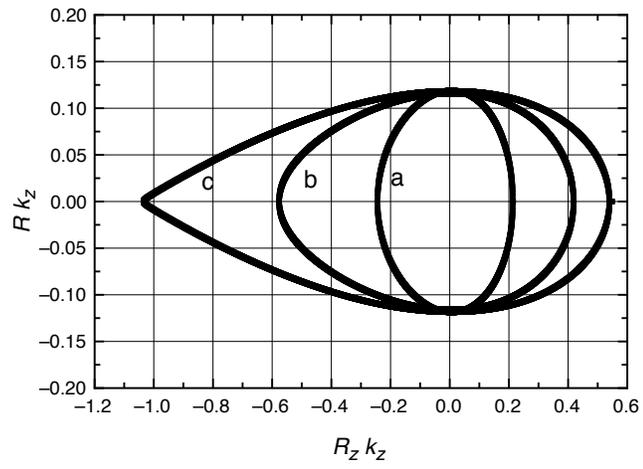

Figure 5: Matched beam profile, Eq. (5.19) for $\varphi_s = -30^o$:
(a) $k_z R_z / |\varphi_s| = 0.38$, (b) $k_z R_z / |\varphi_s| = 0.8$, (c) $k_z R_z / |\varphi_s| = 1$.

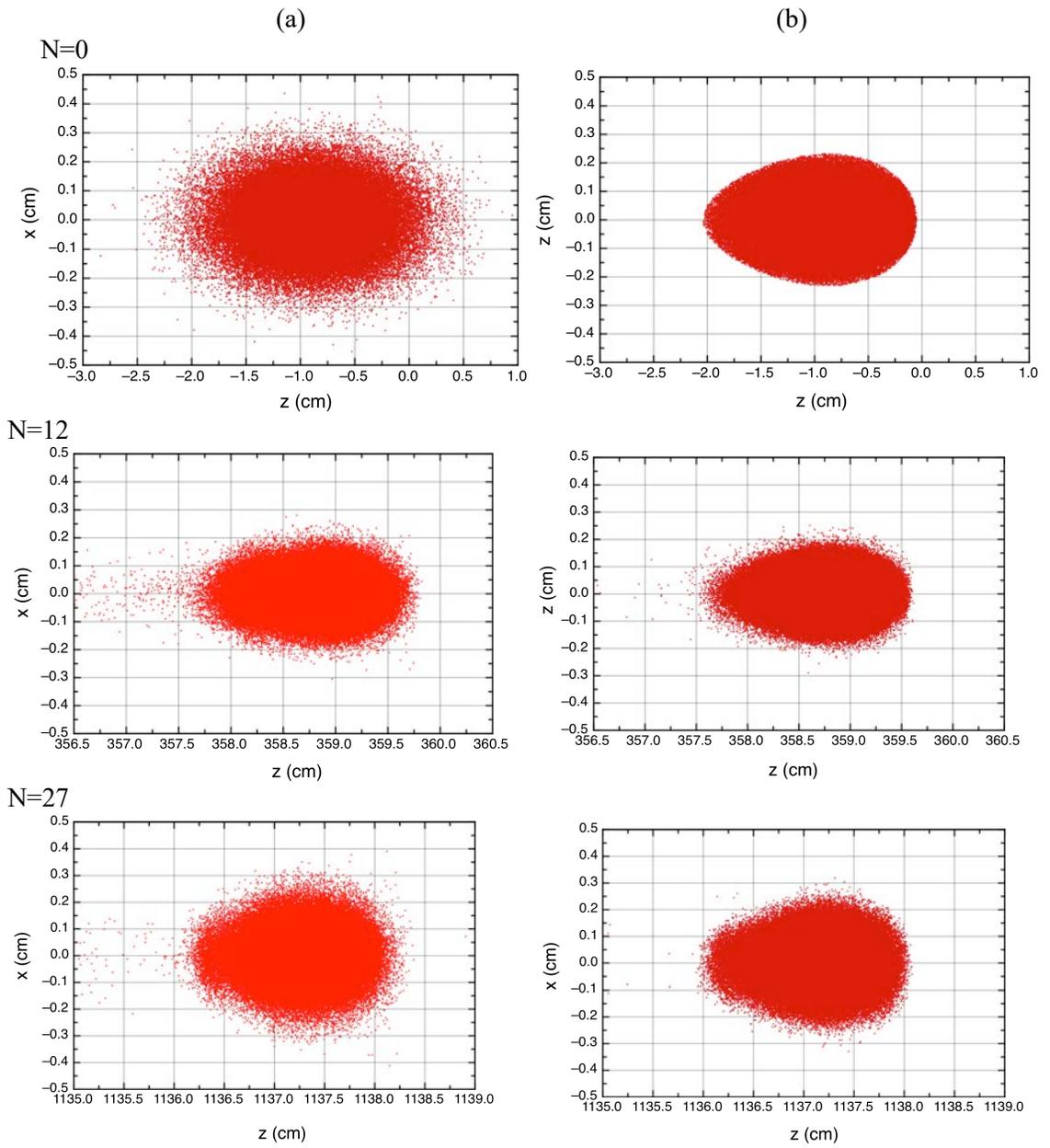

Figure 6: Dynamics of the beam: (left) with initial ellipsoidal shape, (right) with initial matched profile, Eq. (5.19) versus number of focusing periods $N$.

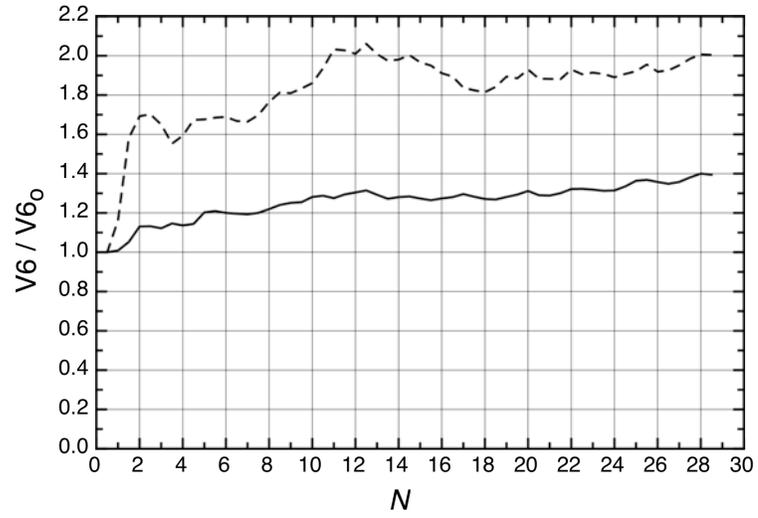

Figure 7: 6D phase space volume growth of the beam: (dotted line) with initial ellipsoidal shape, (solid line) with initial matched profile, Eq. (5.19).

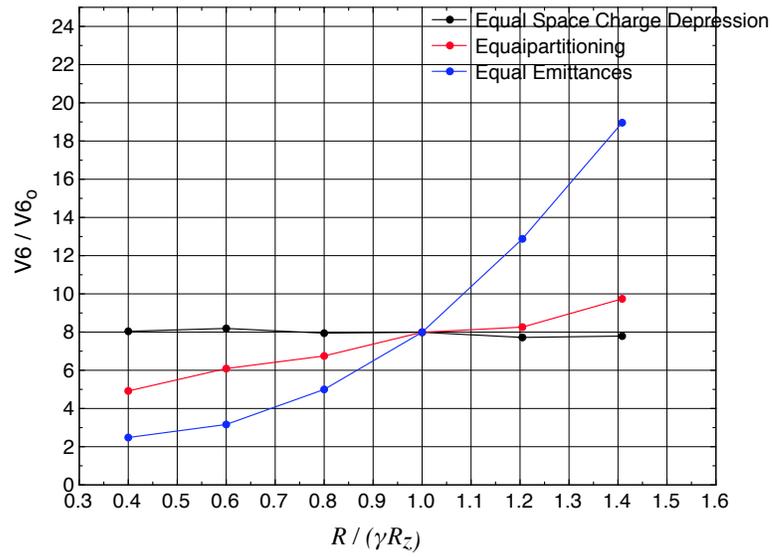

Figure 8: Growth of six-dimensional phase space.

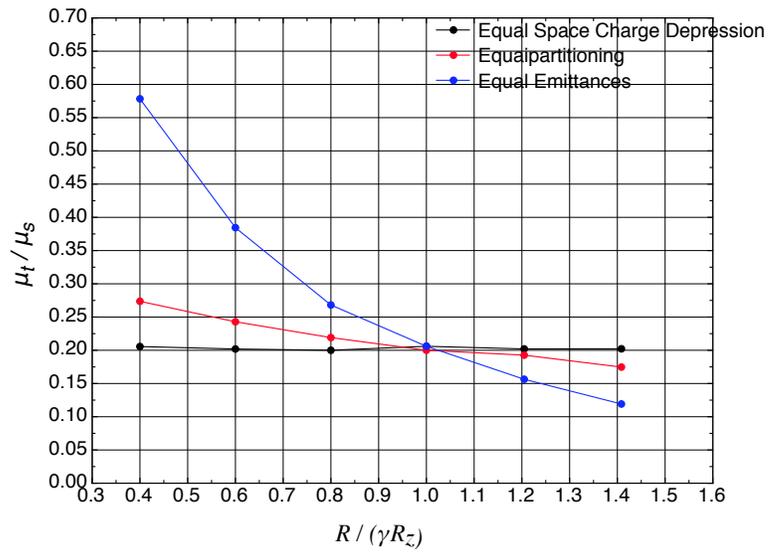

Figure 9: Transverse space charge depression factor versus ratio of beam semi-axis with constant longitudinal space charge depression $\mu_z / \mu_{zo} = 0.2$.

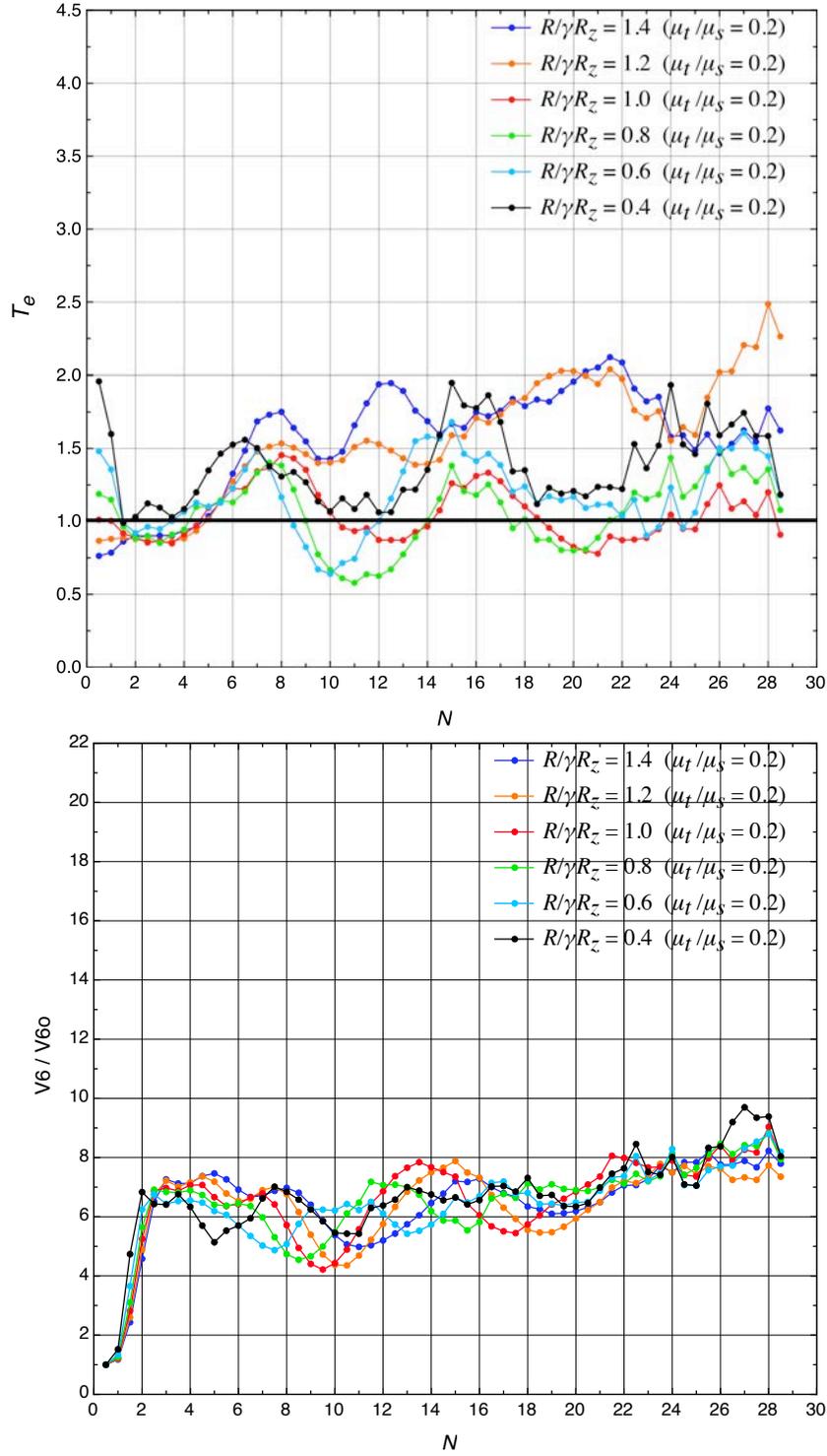

Figure 10: (Up) equipartitioning parameter $T_e$, Eq. (6.7), and (bottom) 6D phase space volume growth as a function of focusing period for equal space charge depression mode.

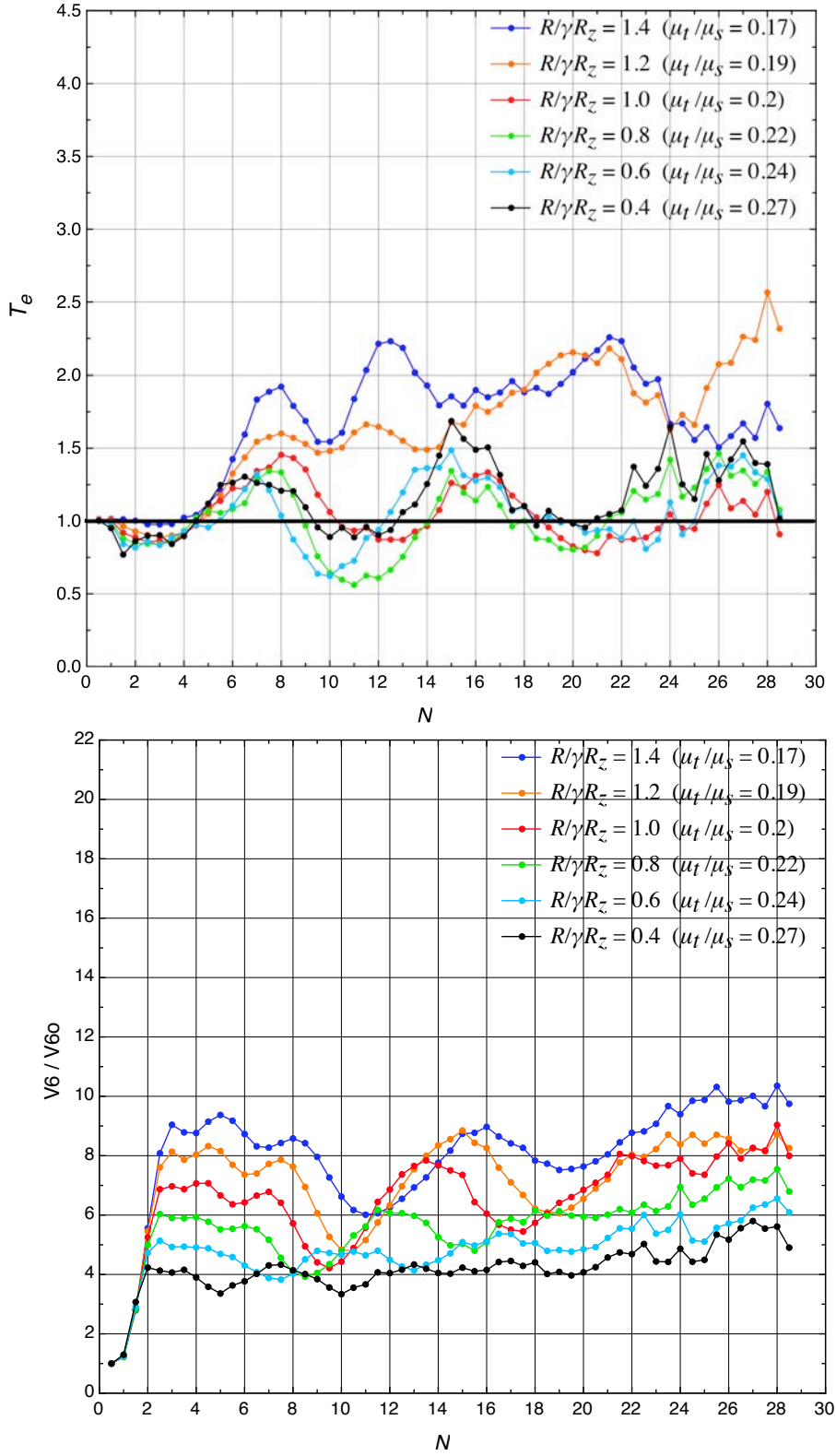

Figure 11: (Up) equipartitioning parameter $T_e$, Eq. (6.7), and (bottom) 6D phase space volume growth as a function of focusing period for equipartitioning mode.

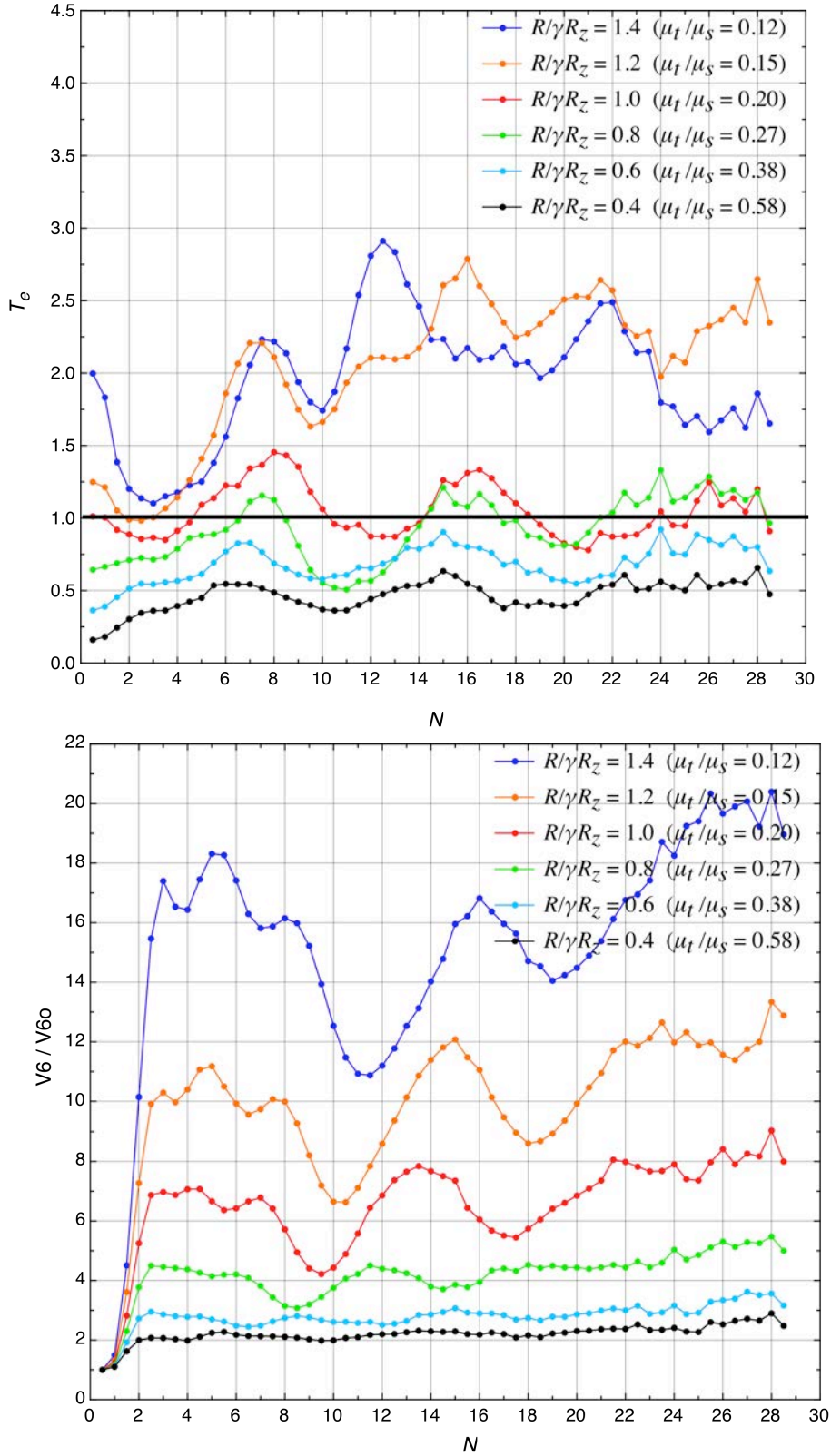

Figure 12: (Up) equipartitioning parameter $T_e$, Eq. (6.7), and (bottom) 6D phase space volume growth as a function of focusing period for equal emittance mode.

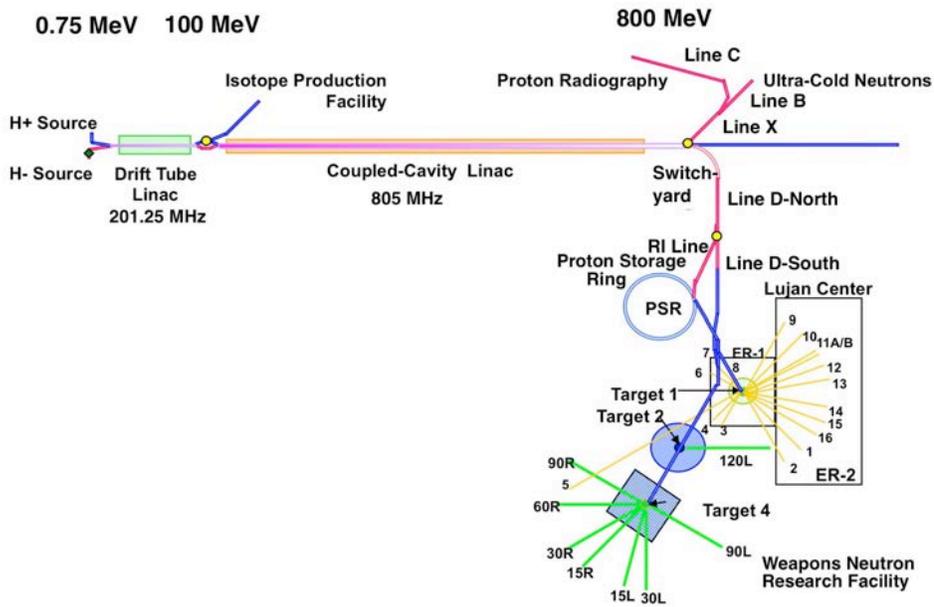

Figure 13: Layout of LANSCE Accelerator Facility.

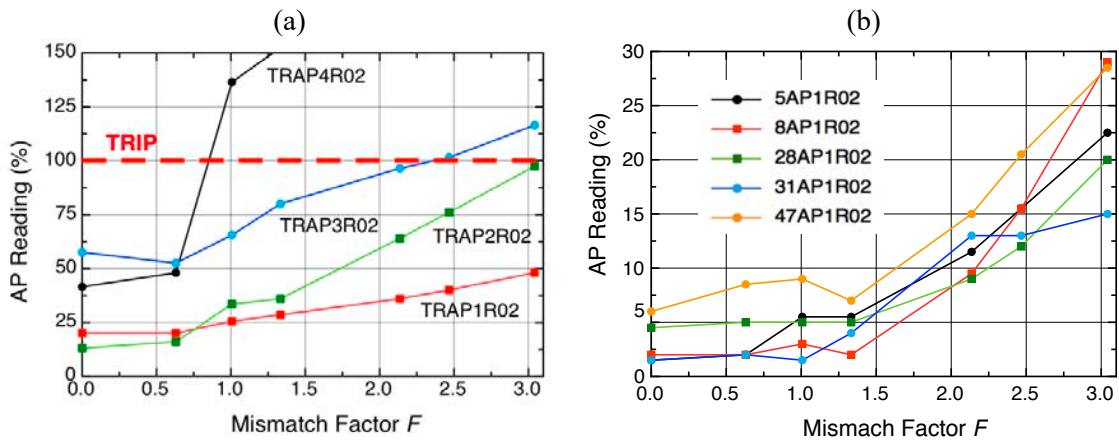

Figure 14: Effect of beam mismatch at the entrance of Drift Tube Linac on beam loss in (a) 100 MeV Transition Region and (b) along Coupled-Cavity Linac.